\documentclass[useAMS,usenatbib]{mn2e}
\usepackage{epsfig}
\usepackage{amsmath}
\usepackage{rotating}
\usepackage{multirow}
\usepackage{natbib}

\newcommand{\be}{\begin{equation}}
\newcommand{\beq}{\begin{equation}}
\newcommand{\ba}{\begin{eqnarray}}
\newcommand{\ee}{\end{equation}}
\newcommand{\eeq}{\end{equation}}
\newcommand{\ea}{\end{eqnarray}}
\newcommand{\msun}{\rm M_{\odot}}

\newcommand{\brunt}{Brunt-V\"ais\"al\"a \,}

\def\lsim{~\rlap{$<$}{\lower 1.0ex\hbox{$\sim$}}}

\def\gsim{~\rlap{$>$}{\lower 1.0ex\hbox{$\sim$}}}

\begin{document} 

\title[Probing Gas Motions in the Intra-Cluster Medium: A Mixture Model Approach]{Probing Gas Motions in the Intra-Cluster Medium: A Mixture Model Approach}
\author[C. Shang and S. P. Oh]{Cien Shang$^{1}$ and S. Peng Oh$^{2}$\\
$^{1}$Kavli Insitute for Theoretical Physics, University of California, Santa Barbara, California, CA 93106; cshang@kitp.ucsb.edu\\
$^{2}$Department of Physics, University of California, Santa Barbara, California, CA 93106; peng@physics.ucsb.edu}

\date{\today}

\maketitle

\label{firstpage}

\begin{abstract}
Upcoming high spectral resolution telescopes, particularly Astro-H, are expected to finally deliver firm quantitative constraints on turbulence in the intra-cluster medium (ICM). We develop a new spectral analysis technique which exploits not just the line width but the entire line shape, and show how the excellent spectral resolution of Astro-H can overcome its relatively poor spatial resolution 
in making detailed inferences about the velocity field. The spectrum is decomposed into distinct components, which can be quantitatively analyzed using Gaussian mixture models. For instance, bulk flows and sloshing produce components with offset means, while partial volume-filling turbulence from AGN or galaxy stirring leads to components with different widths. The offset between components allows us to measure gas bulk motions and separate them from small-scale turbulence, while component fractions and widths constrain the emission weighted volume and turbulent energy density in each component. We apply mixture modeling to a series of analytic toy models as well as numerical simulations of clusters with cold fronts and AGN feedback respectively. From Markov Chain Monte Carlo and Fisher matrix estimates which include line blending and continuum contamination, we show that the mixture parameters can be accurately constrained with Astro-H spectra: at a $\sim 10\%$ level when components differ significantly in width, and a $\sim 1\%$ level when they differ significantly in mean value. We also study error scalings and use information criteria to determine when a mixture model is preferred. Mixture modeling of spectra is a powerful technique which is potentially applicable to other astrophysical scenarios.

%
%
\end{abstract} 

\section{Introduction}
\label{sec:introduction}

Turbulence in galaxy clusters can arise from many sources, ranging from mergers and cosmological structure formation \citep{lau09,vazza09,vazza11,zuhone10}, to galactic wakes \citep{kim07,ruszkowski11}, and AGN feedback (\citet{mcnamara07}, and references therein). It is generally expected to be highly subsonic, with Mach numbers ${\cal M}\sim 0.1-0.5$. Turbulence has wide-ranging and pivotal effects on ICM physics. It could dominate metal transport \citep{rebusco05,simionescu08}, accelerate particles, as required in a prominent model of radio halos \citep{brunetti01,brunetti07}, generate and amplify magnetic fields \citep{subramanian06,ryu08,cho09,ruszkowski11a}, and provide pressure support, thus impacting X-ray mass measurements \citep{lau09}, and Sunyaev-Zeldovich (SZ) measurements of the thermal pressure \citep{shaw10,battaglia11,battaglia11a,parrish12}. The unknown level of non-thermal pressure support introduces systematic deviations in the mass calibration of clusters and could strongly affect their use for cosmology. A particularly interesting effect of turbulence is its impact on the thermal state of the gas, potentially allowing it to stave off catastrophic cooling. It can do by dissipation of turbulent motions \citep{churazov04,kunz11}, or turbulent diffusion of heat \citep{cho03,kim03,dennis05}. More subtly, it can do so by affecting magnetic field topology; by randomizing the $B$-field, it can restore thermal conduction to $\sim 1/3$ of the Spitzer rate \citep{ruszkowski10,ruszkowski11,parrish10}. Besides turbulence, a variety of bulk motions such as streaming, shocks, and sloshing have been observed\footnote{While rotation has not been directly seen, it is also expected from cosmological simulations \citep{lau11}. Its effects are generally too small to be detected by the methods discussed in this paper.}. Such (often laminar) gas motions are interesting in their own right. For instance, gas sloshing in the potential well of clusters, which produces observed cold fronts---contact discontinuities between gas of very different entropies---has gleaned information about hydrodynamic instabilities, magnetic fields,
thermal conductivity and viscosity of ICM \citep[and references
therein]{Markevitch2007}.

Current observational constraints on ICM turbulence are fairly weak, and mostly indirect. They come from the analysis of pressure maps \citep{schuecker04}, the lack of detection of resonant-line scattering \citep{churazov04,werner10}, Faraday rotation maps \citep{vogt05,enslin06}, and deviations from hydrostatic equilibrium with thermal pressure alone \citep{Churazov2008,churazov10,Zhang2008}. In general, these studies constrain cluster cores and either place upper bounds on turbulence, or indicate (with large uncertainties) that it could be present with energy densities $\sim 5-30\%$ that of thermal values. The energy density in turbulence is expected to increase strongly with radius \citep{shaw10,battaglia11}, though observational evidence for this is indirect. The most direct means to constrain gas motions is through Doppler broadening of strong emission lines, but this remains undetected with current technology. By examining the widths of emission lines with XMM RGS,
\citet{Sanders2010} found a 90\% upper limit of 274 ${\rm km s^{-1}}$
(13\% of the sound speed) on the turbulent velocity in the inner 30
kpc of Abell 1385; analysis of other systems provides much weaker bounds ($\lsim 500 {\rm km s^{-1}}$; \citet{sanders11}). 
Numerical simulations provide further insights
\citep[e.g.][]{Lau2009,Vazza2009, Vazza2010}. However, due to limited resolution and frequent exclusion of important physical
ingredients such as AGN jets, magnetic fields, radiative cooling, anisotropic viscosity, it is difficult to draw robust conclusions. 

The forthcoming  Astro-H mission\footnote{http://astro-h.isas.jaxa.jp/} (launch date 2014)
represents our best hope of gaining a robust understanding of gas motions in 
ICM\footnote{Much farther in the future, the ATHENA mission (http://sci.esa.int/ixo) could significantly advance the same goals.}. With unprecedented spectral resolution (FWHM $\sim$ 4-5 eV), 
Astro-H could not only measure the widths of emission lines,
therefore constraining the turbulent amplitude, but also probe the line
shapes. Somewhat surprisingly, very few studies have been conducted to extract velocity
information from the shape of emission lines in a realistic observational context. Current work has focused on studying a Gaussian approximation to the line \citep{rebusco08}, and interpreting the radial variation of the line width and line center \citep{zhuravleva12}. For instance, \citet{zhuravleva12} show how the radial variation of line width is related to the structure function of the velocity field, and also how the 3D velocity field can be recovered from the projected velocity field. However, such inferences generally require angular resolution comparable to characteristic scale lengths of the velocity field, and are likely feasible only for one or two very nearby clusters such as Perseus (though such studies do represent a very exciting possibility for ATHENA). At the same time, it has long been apparent that turbulence in clusters leads to significant non-Gaussianity in the line shape---indeed, these were clearly visible in the early simulations of \cite{sunyaev03}. \citet{Inogamov2003} presented a deep and insightful discussion of the origin of line shapes, albeit in an idealized Kolmogorov cascade model for cluster turbulence (for instance, they do not consider the effect of gas sloshing and cold fronts). Heuristically, one can consider non-Gaussianity to arise when the size of the emitting region (heavily weighted toward the center in clusters) is not much larger than the characteristic outer scale of the velocity field. The central limit theorem does not hold as the number of independent emitters is small (and/or in large scale bulk flows, the motion of different emitters is highly correlated). In a series of papers, Lazarian and his
collaborators considered the relationship between the turbulent
spectrum and the spectral line shape in the ISM \citep{Lazarian2000,
  Lazarian2006,Chepurnov2006}. However, since they focused on the supersonic and compressible turbulence seen in the ISM---a regime where thermal broadening is negligible and density fluctuations are considerable---the methods they employ are not readily suitable for the mild subsonic turbulence expected in the ICM. 

We therefore aim to study how velocity information 
can be recovered from the emission line profile in the ICM context, in a realistic observational setting. In particular, we advance the notion that the profile can be separated into different
modes, which have a meaningful physical interpretation. As we will discuss in more detail below, many processes in the
ICM could give rise to velocity fields composed of
distinct components. For instance, the sharp contact discontinuity in velocity in cold fronts will give rise to a bimodal velocity field where one component is significantly offset from another. Another interesting scenario arises if turbulence is not volume-filling (due, for instance, to anistropic stirring by AGN jets). Then spectral lines of different width (with and without turbulent broadening) will be superimposed on one another. When seen in the same field of view (FOV),
these components correspond to different modes in the line
profile, and decomposing the velocity field into dominant modes can
yield valuable quantitative information (for instance, the volume
filling factor). For the upcoming Astro-H mission, mode
separation in the spectrum is necessary and important since the poor
angular resolution make it hard to spatially resolve different
components---indeed, the high spectral resolution of Astro-H is our
best tool for inferring the complex structure of the velocity
field. We use standard mixture modeling techniques and Fisher
matrix/Markov chain Monte Carlo error analysis to quantify how well we
could separate and constrain different 
components from a single spectrum, and then establish what we can learn from about the underlying velocity field from such a component separation. 

\begin{table}
  \caption{Specifications of the Soft X-ray Spectroscopy System
    onboard the Astro-H telescope.}
    \label{tbl:specs}
\begin{center}
\begin{tabular}{l c}
\hline \hline
Effective area ${\rm cm^2}$ at 6 keV& 225\\
Energy range (keV) & 0.3-12.0  \\
Angular resolution in half power diameter (arcmin) & 1.3\\
Field of view (${\rm arcmin}^2)$ & $3.05\times 3.05$\\
Energy resolution in FWHM (eV) & 5\\
\hline
\end{tabular}
\end{center}
\end{table}

\begin{table}
  \caption{Photon counts $N_p$ in the He-like iron line and physical length
corresponding to angular resolution in HPD (1.3 arcmin) for a few
nearby galaxy clusters. The photons are accumulated from 1 FOV through
the cluster center over $10^6$ seconds. }
    \label{tbl:clusters}
\begin{center}
\begin{tabular}{l c c c }
\hline \hline
Cluster Name& Redshift & $d_{1.3}$ (kpc) & $N_p (\times
10^4~{\rm phot})$ \\
\hline
PERSEUS & 0.0183 & 28.81 & 5.8\\
PKS0745 & 0.1028 & 146.76 & 3.8\\
A0478 & 0.0900 & 130.36 & 3.7\\
A2029 & 0.0767 & 112.79 & 3.4\\
A0085 & 0.0556 & 83.78 & 3.1\\
A1795 & 0.0616 & 92.17 & 2.6\\
A0496 & 0.0328 & 50.76 & 1.9\\
A3571 & 0.0397 & 60.94 & 1.7\\
A3112 & 0.0750 & 110.51 & 1.7\\
A2142 & 0.0899 & 130.23 & 1.6\\
2A0335 & 0.0349 & 53.87 & 1.3\\
HYDRA-A & 0.0538 & 81.23 & 1.3\\
A1651 & 0.0860 & 125.13 & 1.1\\
A3526 & 0.0103 & 16.37 & 0.8\\
\hline
\hline
\end{tabular}
\end{center}
\end{table}

Before proceeding to the main discussion, we first list a few
specifications of the Astro-H mission, on which our
discussions are based. Our study mainly takes the advantage of the
high spectral resolution of the Soft X-ray Spectroscopy System (SXS)
onboard the Astro-H telescope. Its properties, taken from the
``Astro-H Quick
Reference''\footnote{http://astro-h.isas.jaxa.jp/doc/ahqr.pdf}, are
given in Table \ref{tbl:specs}. 
The energy resolution is 5 eV in FWHM\footnote{It has shown to be even lower--4 eV--in laboratory tests \citep{porter10}.},
corresponding to a standard deviation of 2.12 eV. For comparison, the
thermal broadening of the Fe 6.7 keV line is 2.07 eV for a 5 keV cluster, while
broadening by isotropic Mach number ${\cal M} \sim 0.2$ motions is 2.9 eV. Thus, for the highly subsonic motions in the core with Mach numbers ${\cal M} \sim 0.1-0.3$ generally seen in cosmological simulations, the instrumental, thermal and turbulent contributions to line broadening are all roughly comparable. In contrast to the impressive energy resolution, the angular
resolution of Astro-H is poor: 1.3 arcmin in half power diameter
(HPD). 
Therefore, different velocity components are likely to show up in
the same spectrum. Based on these specifications, Table
\ref{tbl:clusters} shows the expected photon counts in the He-like
iron line at 6.7 keV for a few of the brightest nearby clusters ($z \leq 0.1$). The photons are accumulated in one FOV through
the cluster center over $10^6$ seconds; the $\sim$ several x $10^{4}$ photons collected should allow good statistical separation of mixtures if present.  The density distributions and
cluster temperatures are taken from \citet{Chen2007}, and metallicity
is assumed to be 0.3 ${\rm Z_{\odot}}$. 
Also shown are the physical lengths corresponding to the
angular resolution in HPD, which are 
$\sim$100 kpc; comparable to the core size. We therefore do not expect Astro-H to spatially resolve many
structures.

The remainder of the paper is organized as follows. In
\S~\ref{sec:motivations}, we discuss possible scenarios that could
give rise to multiple component spectra, further motivating the current 
study. In \S~\ref{sec:methodology}, we develop the methodology to be
used in this paper. In \S~\ref{sec:constraints}, we discuss how
accurately different components could be recovered in idealized toy models, to build our understanding of the applicability and capabilities of the method. In \S~\ref{sec:application}, we
apply our statistical method to realistic simulations of galaxy clusters, where we have full knowledge of the underlying velocity field, and see what information we can recover. In \S~\ref{sec:conclusions}, we
conclude by summarizing the main results.

\section{Motivation}
\label{sec:motivations}

In this section, we motivate the current study by giving examples of very common processes operating in the ICM which could give rise to multi-component velocity fields: bulk motions from mergers and sloshing, and AGN feedback. 

\subsection{Bulk Motions}
\label{subsec:example}

\begin{figure}
\begin{tabular}{c}
\rotatebox{-0}{\resizebox{120mm}{!}{\includegraphics{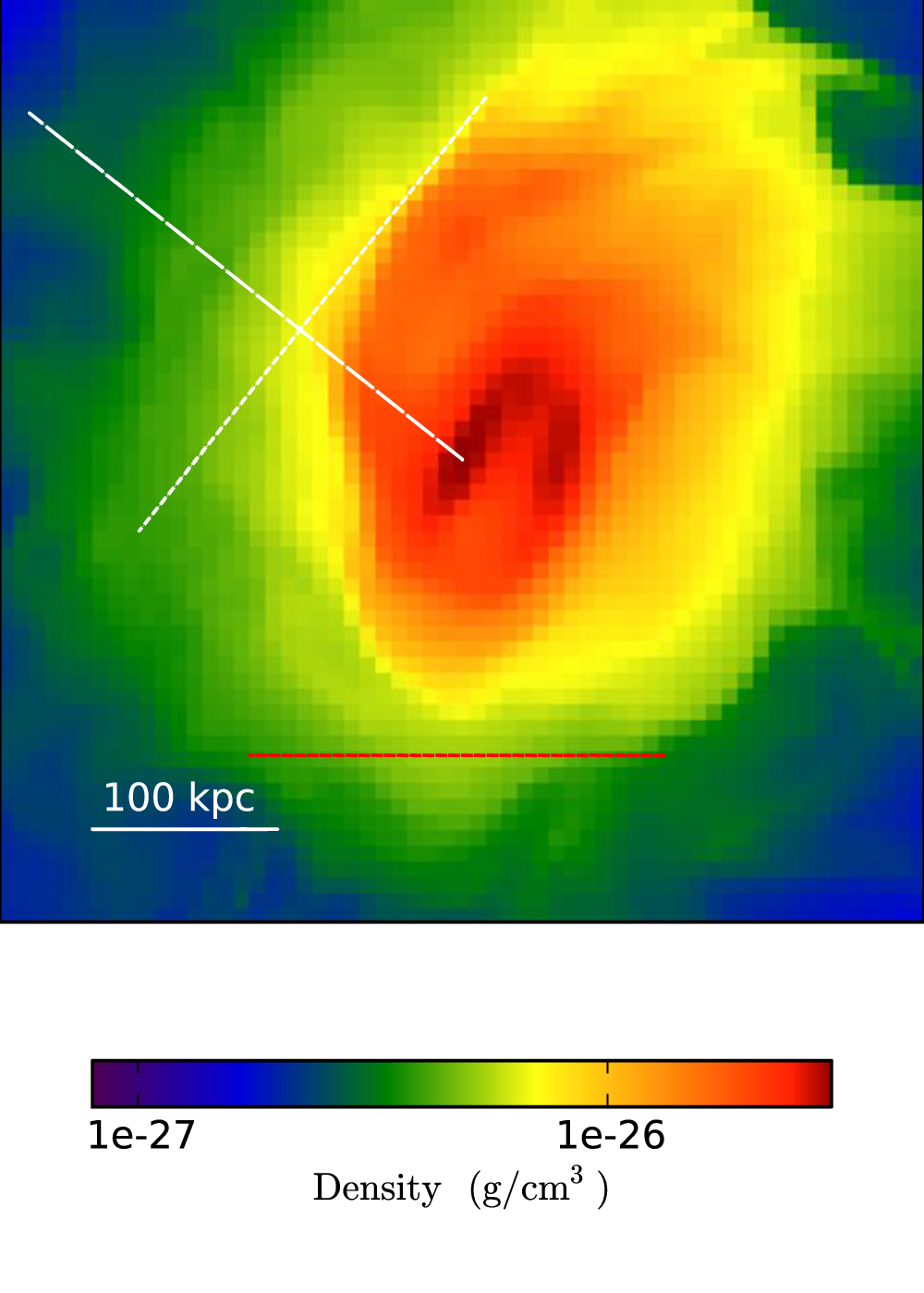}}}
\end{tabular}
\caption{Density map on a slice through the cluster center. The dashed line shows the direction along which the profiles in Fig. \ref{fig:profiles} are
computed, while the perpendicular dotted line--chosen to maximize line of sight velocity shear
--indicates the observation direction for the solid red velocity PDF shown in
Fig. \ref{fig:spectrum}. The dotted red line shows an alternate viewing direction with much less velocity shear; its velocity PDF is given by the thin red line in Fig. \ref{fig:profiles}.} 
\label{fig:density}
\end{figure}

\begin{figure}
\begin{tabular}{c}
\rotatebox{-0}{\resizebox{90mm}{!}{\includegraphics{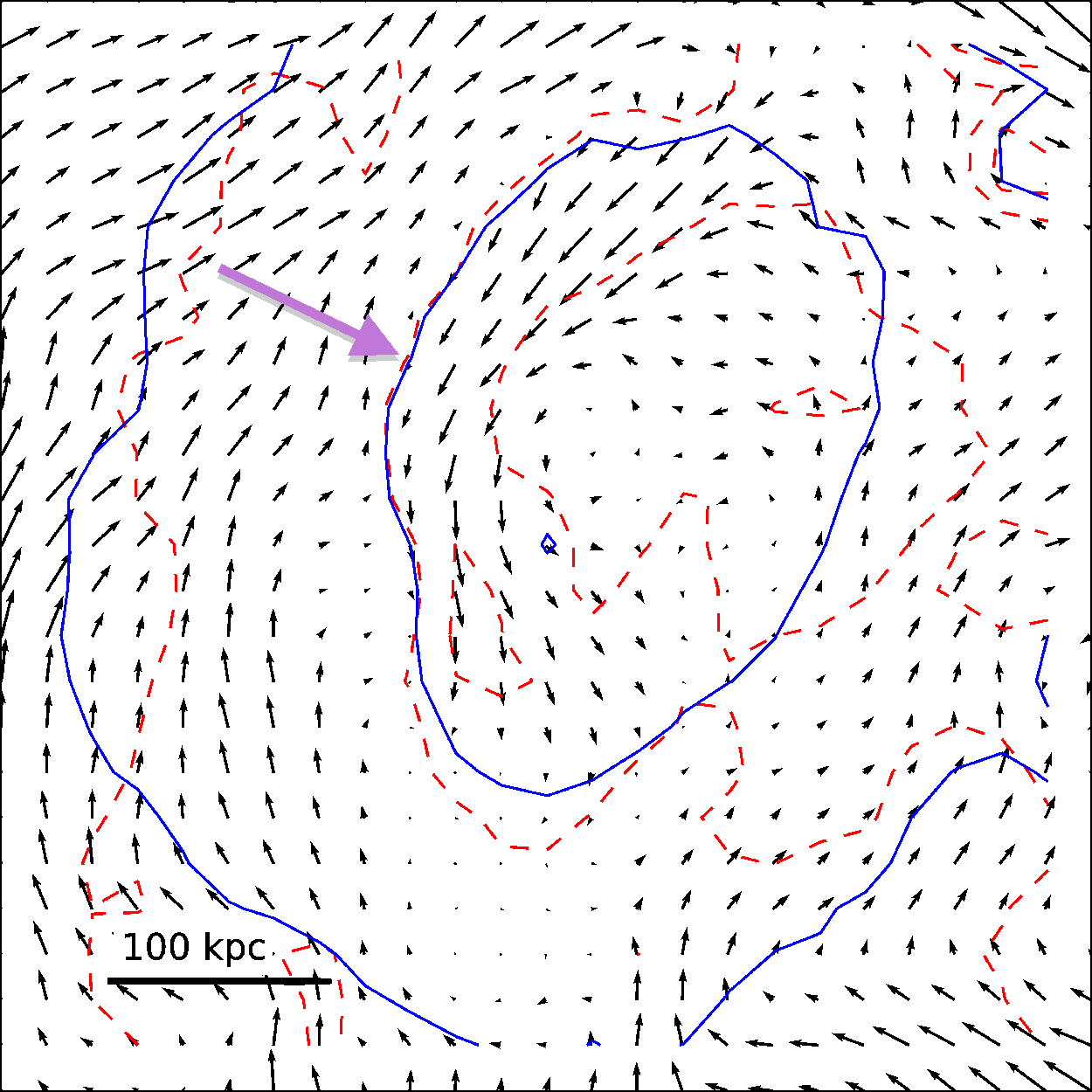}}}
\end{tabular}
\caption{Velocity fields on the same slide as in Fig.
  \ref{fig:density}, overlaid with density (solid blue curves) and
  temperature (dashed red curves) contours. The large purple arrow
  indicates the location of the cold front.}
\label{fig:velocity}
\end{figure}

\begin{figure}
\begin{tabular}{c}
\rotatebox{-90}{\resizebox{90mm}{!}{\includegraphics{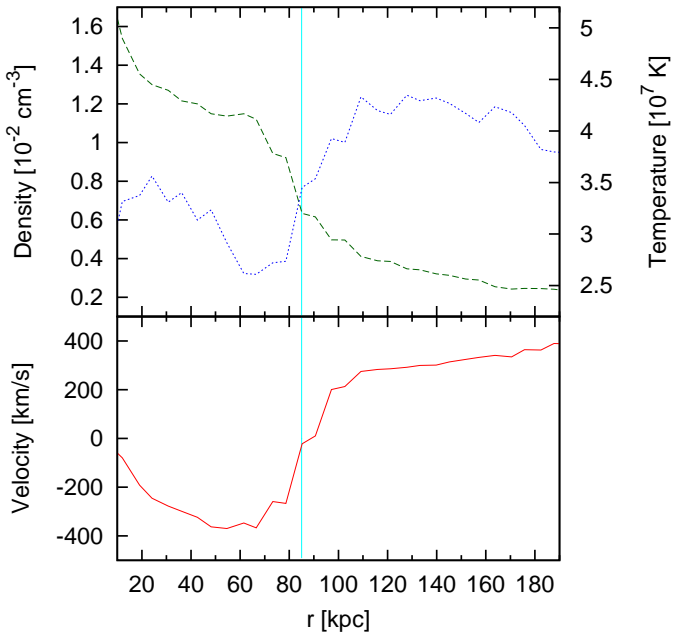}}}
\end{tabular}
\caption{Density (dashed curve), temperature (dotted curve)  and
  line-of-sight velocity (solid curve in the bottom panel) profiles
  along the a direction perpendicular to the cold front, as
  indicated in Fig. \ref{fig:density} with a dashed line. 
  Here, the position of the cold front is given by the
  vertical (cyan) line at 85 kpc. }
\label{fig:profiles}
\end{figure}

\begin{figure}
\begin{tabular}{c}
\rotatebox{-90}{\resizebox{60mm}{!}{\includegraphics{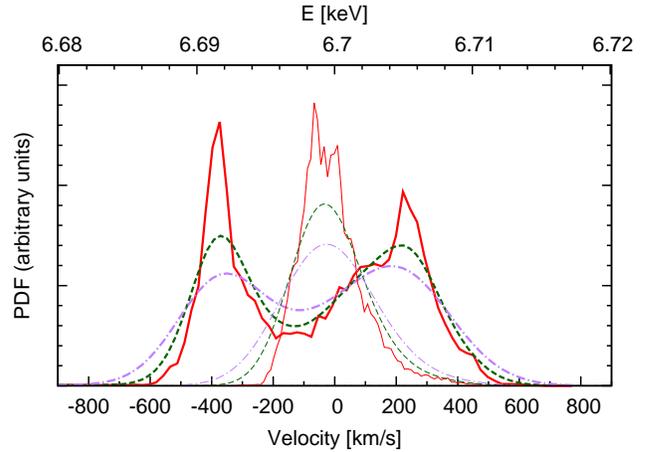}}}
\end{tabular}
\caption{The thick solid (red) curve is the normalized emission-weighted
  velocity PDF from a box centered on the dotted line in Fig.
  \ref{fig:density}. The box is 100 kpc long, 100 kpc wide and 1 Mpc
  deep. The thick dashed green curve,
  is the corresponding profile of the He-like iron line at 6.7 keV
  (see top axis for energy scale), including the effects of thermal
  broadening, while the thick dot-dashed purple curve also includes
  instrumental broadening.
The thin lines show the same curves for
the line of sight given by the dotted red line in
Fig. \ref{fig:density}.
}
\label{fig:spectrum}
\end{figure}

Thus far, most constraints on gas bulk motions comes from observations of sharp density gradients in the plane of the sky. Classic bow shocks have been seen in a handful of violent mergers. Much more common are ``cold fronts''  \citep{markevitch07}: sharp contact discontinuities between gas phases of different entropies, discovered in the last decade thanks to the high-resolution of the {\it Chandra} X-ray telescope. They are seen both in mergers (where the cold gas arises from the surviving cores of infalling subclusters) and relaxed cool core clusters (where they are produced by the displacement and subsequent sloshing of the low-entropy central gas in the gravitational potential well of the cluster). They are remarkably ubiquitous, even in relaxed cool core clusters with no signs of recent mergers, which often exhibit several such cold fronts at different radii from the density peak. For instance, they are seen in more than half of all cool core clusters; given projection effects, most if not all cool core clusters should exhibit such features. Evidently, coherent gas bulk motions are extremely common if not universal\footnote{Indeed, we show here the very first cluster we simulated from random initial conditions, which already exhibited cold front like features.}, and their effects must be taken into account when interpreting Astro-H spectra. Generically, we would expect bulk motions to offset the centroids of emitting regions with significant line-of-sight relative velocity. Cold fronts have been used to probe the amplitude and direction of gas motions in the plane of the sky; combining this with line-of-sight information from the spectrum could prove very powerful indeed. 

Our example is taken from an adiabatic numerical simulation from cosmological initial conditions with the
adaptive mesh code Enzo \citep{Bryan1999,Norman1999,O'Shea2004}. 
We assume a $\Lambda$CDM cosmology with cosmological parameters
consistent with the seventh year {\it WMAP} results \citep{komatsu11}:
$\Omega_m=0.274$, $\Omega_{\Lambda}=0.726$, $\Omega_b=0.045$,
$h=0.705$, $\sigma_8=0.810$, $n_s=0.96$. 
The
simulation has a box size of 64 
Mpc, and a root grid of $128^3$. We picked the most massive cluster
($M \sim 2\times 10^{14}~\msun$) from the fixed-grid initial run, and
re-simulate it with much higher resolution. The highest spatial resolution is
11 kpc in the cluster center. 

The cluster has a disturbed morphology, and shows a ``cold
front''-like feature in the core. 
Note that our adiabatic simulation necessarily produces a NCC cluster. 
The density and velocity fields on a
slice through the cluster center are shown in Fig. \ref{fig:density} and \ref{fig:velocity},
respectively. In the position indicated by the large arrow in Fig.
\ref{fig:velocity}, the density, temperature and velocity all change
rapidly. This is clearly shown in Fig. \ref{fig:profiles}, which shows
density, temperature and velocity profiles along a line perpendicular
to the front (indicated in Fig. \ref{fig:density} with a dashed
line). At $\sim 85$ kpc from the cluster center, the density decreases
while the temperature increases rapidly, as expected in a cold
front (for a shock, the temperature jump would be opposite). Furthermore, the pressure is continuous across the front, 
while the tangential velocity changes direction discontinuously across the front---both well-known features of cold fronts \citep{markevitch07}. 

For an observation direction along the white dotted line in Fig.
\ref{fig:density}, Fig. \ref{fig:spectrum} shows the
emission-weighted probability distribution function (PDF) of the
line-of-sight velocity. Motivated by Table \ref{tbl:clusters}, we
extract the emission-weighted PDF from an volume with an area of $100\times100~{\rm kpc}^2$ and a
depth of 1 Mpc (this last number represents the line of sight depth, and is chosen for convenience. Our results are insensitive to it as long as it is much larger than the core size, where most of the photons come from). 
The PDF clearly shows two peaks, centered at -400 ${\rm km \, s^{-1}}$ 
and 250 ${\rm km \, s^{-1}}$, corresponding to the gas on different side of the
cold front. After convolution with thermal broadening,
the dashed line shows the profile of the He-like iron line at 6.7 keV,
while the dot-dashed line also includes the instrumental broadening of
Astro-H. They also clearly 
show double peak features.

The above case is a somewhat idealized ``best case'' scenario, where we have assumed the
viewing angle to be along the direction of maximum line of sight velocity shear, thus 
maximizing the separation between the two peaks in the velocity
PDF. For a more general viewing angle, the separation would not be so
clear, as we show with the thin curves in
Fig. \ref{fig:spectrum}. This is the PDF along the red dotted curve in
\ref{fig:density},
which has very small line-of-sight bulk flow. 
There is only one large peak, but with a long tail. From 
Fig. \ref{fig:velocity}, we see this long tail comes from the
gas surrounding the cold clump, which has shear 
velocities with large components along the LOS. 
Therefore the PDF can also be separated into two
components -- a narrow component emitted by the cold clump and a broad
component from the ambient gas. The offset between the components is a measure of the LOS contact discontinuity in the bulk velocity, while smaller scale shear contributes to the width. 
Such a decomposition of the line-of-sight velocity, combined with spatially resolved temperature and density information in the plane of the sky from X-ray imaging, could shed more light on the 3D velocity field as well as physical information such as the gas viscosity. 

\subsection{Volume-filling Factor of Turbulence}
\label{subsec:others}

The previous section highlighted a situation where strong shear or bulk
motion gives rise to different components with offset centroids
(``separation driven'' case). Another regime where different components
could arise is when the two components have markedly different widths
(``width driven'' case). We saw an example of this at the end of the
previous section: a narrow component due to a cold, kinematically
quiescent clump, and a broader component due to the sheared
surrounding ambient gas. More generally, different widths arise when
turbulence varies spatially. The case when turbulence is only
partially volume-filling is a particularly interesting special
case. Many of the physical effects of turbulence depend not only on
its energy density, but its volume filling fraction $f_{\rm V}$, which
is often implicitly assumed to be unity. For instance, for turbulence
to stave off catastrophic cooling, it must be volume-filling. This is
by no means assured. For instance, analytic models
\citep{subramanian06} of turbulence generation during minor mergers
predict $f_{\rm V}\sim 0.2-0.3$ to be small, but area-filling (i.e.,
the projection of turbulent wakes on the sky cover a large fraction of
the cluster area, $f_{\rm S} \sim O(1)$). Interestingly, cosmological
AMR simulations which use vorticity as a diagnostic for turbulence
find good agreement; $f_{\rm V} \lsim 0.3$ and $f_{\rm S} \sim O(1)$
for all runs \citep{iapichino08}. In our own simulations of stirring
by galaxies \citep{ruszkowski11}, we have seen both high and low
values of $f_{\rm V}$, depending on modeling assumptions. If $g$ modes
are excited by orbiting galaxies (which requires the driving orbital
frequency $\omega$ to be less than the \brunt frequency $\omega_{\rm
  BV}$--a requirement which depends on both the gravitational
potential and temperature/entropy profile of the gas), then
volume-filling turbulence is excited; otherwise turbulence excited by
dynamical friction is potentially confined to thin ``streaks'' behind
galaxies (see also \citet{balbus90,kim07}). If turbulence is patchy,
we might expect spectral lines to have a narrow thermal Doppler core
(produced in quiescent regions), with turbulently broadened tails. In
the context of our mixture model, measuring the fraction of photons in
the second component might allow a quantitive measure of $f_{\rm V}$.

Yet another context in which strongly spatially varying or partial volume-filling turbulence could result in multiple components in the velocity PDF is AGN feedback, which is ubiquitous in most cool core clusters (\citet{Birzan2004}; for a recent review, see \citet{mcnamara12}). AGN jets are launched over a narrow solid angle and are fundamentally anisotropic; thus, their ability to sustain isotropic heating in the core has often been questioned. Isotropization of the injected energy could arise from weak shocks and sound waves \citep{fabian03}, frequent re-orientation of jets by randomly oriented accretion disks \citep{king07}, jet precession \citep{dunn06,gaspari11}, and cavities being blown about by cluster weather \citep{bruggen05,heinz06,morsony10}. As above, AGN could also excite g-mode oscillations; an intriguing example is a cross-like structure on 100 kpc scales in the ICM surrounding 3C 401 \citep{reynolds05}. A measurement of $f_{\rm V}$ could thus constrain the efficacy of such mechanisms in isotropizing AGN energy deposition throughout the core. The expansion of AGN-driven cavities can also introduce high bulk velocities and shear (corresponding more to the ``separation-drive'' regime); this is potentially directly measurable with ATHENA's excellent angular and spectral resolution \citep{Heinz2010a}, but would require indirect methods such as mixture modeling with the poor angular resolution of Astro-H. In \S \ref{sec:application}, we analyze an AGN feedback simulation kindly provided to us by M. Br\"{u}ggen. 

\subsection{Physical Significance of Mixture Model Parameters}
\label{section:physical_significance} 

In \S \ref{sec:methodology}, we lay out our methodology for recovering mixture model parameters, and in subsequent sections we describe how accurately these can be constrained. These parameters are the mixture weights $f_{i}$, and means and variances $\mu_{i}, \sigma_{i}^{2}$, of the fitted Gaussians. Given the results of this section, we can tentatively ascribe physical significance to these parameters. The mixture weights $f_{i}$ represent the emission-weighted fraction of the volume in each distinguishable velocity component. The Gaussian means $\mu_{i}$ represent the bulk velocity of a given component. In particular, the difference between the means is a measure of the LOS shear between these components (e.g., as arises at a cold front). Note that this shear due to bulk motions can be considerably larger than the centroid shift due to variance in the mean, induced by turbulent motion with a finite coherence length. The latter is given by $\mu_{i} \sim \sigma_{i}/\
\sqrt{N}$, where $N\sim L_{\rm emit}/l_{\rm v}$ is the average number of eddies pierced by the line of sight, and $L_{\rm emit},l_{\rm v}$ are the size of the emitting region and the coherence length of the velocity field respectively \citep{rebusco08,zhuravleva12}. The variances $\sigma_{i}$ represents turbulent broadening or shear due to the small scale velocity field.

\section{Methodology}
\label{sec:methodology}

We have argued that the X-ray spectrum from galaxy clusters should have multiple distinct components. Uncovering these components is the domain of {\it mixture modeling}, a mature field of statistics with a large body of literature. We will specialize to the case of Gaussian mixture modeling, when Gaussians are used as the set of basis functions for the different components. This is an obvious choice, since thermal and instrumental broadening are both Gaussian, and turbulent
broadening can be well approximated with a Gaussian  when the injection scale is much smaller than the size of the emitting regions
(\citet{Inogamov2003}; i.e., once coherent bulk motions have been separated out by classification into different mixtures, the remaining small scale velocity field is well approximated by a Gaussian). It is also by far the best studied case. Mixture modeling has been applied to many problems in astrophysics, such as detecting bimodality in globular cluster metallicities \citep{ashman94, muratov10} linear regression \citep{kelly07}, background-source separation \citep{guglielmetti09}, and detecting variability in time-series \citep{shin09}, though to our knowledge it has not been applied to analyzing spectra. It should be noted that the specialization to Gaussian mixture is not necessarily restrictive; for instance, Gaussian mixtures have been used to model quasar luminosity functions \citep{kelly08}. For us, the fact that Gaussians are a natural basis function allows us to model the spectra compactly with a small number of mixtures, and assign physical meaning to these different components. 

Consider a model in which the observations $x_{1},\ldots,x_{n}$ are distributed as a sum of $k$ Gaussian mixtures: 
\begin{equation}
f(x|\theta) = \sum^{k}_{j=1} \omega_{i} f_{j}(x|\mu_{j},\sigma^{2}_{j}), 
\label{eqn:mixture} 
\end{equation}
where $f_{j}(x|\mu_{j},\sigma^{2}_{j})$ are normal densities with
unknown means $\mu_{j}$ and variances $\sigma_{j}^{2}$, and
$\omega_{i}$ are the mixture weights. The parameters which must be
estimated for each mixture are therefore
$\theta_{j}=(\omega_{j},\mu_{j},\sigma_{j}^{2})$, and the function $f$
can be viewed as the probability of drawing a data point with value
$x$ given the model parameters $\theta$. Parameter estimation in this
case suffers from the well-known {\it missing data problem}, in the
sense that the information on which distribution $j$ a data point
$x_{i}$ belongs to has been lost. In addition, the number of mixtures
$k$ may not be a priori known\footnote{In this case, the optimal
  number of mixtures can also be estimated from the data, via simple
  criteria such as the Bayesian Information Criterion (see equation
  \ref{eqn:bic}), or more sophisticated techniques in so-called
  Infinite Gaussian Mixture Models. In this paper, we only investigate
  separating the two most dominant components of the spectrum, which
  have the highest signal-to-noise. The data is generally not of
  sufficient quality to allow solving for more than two mixtures
  (strong parameter degeneracies develop). Physical interpretation is
  also most straightforward for the two dominant mixtures.}. Standard
techniques for overcoming this are a variant of maximum likelihood
techniques known as Expectation Maximization (EM; \citet{dempster77}),
or Maximum a Posterior estimation (MAP; see references in Appendix),
which generally involves Markov Chain Monte Carlo (MCMC) sampling from
the posterior. They are both two-step iterative procedures in which
parameter estimation and data point membership are considered
separately. Since they do not require binning of the data, all
information is preserved. We have experimented extensively with
both. However, due to the large number of data points ($\sim 10^{4}$
photons) in this application, we have found that the much simpler and
faster procedure of fitting to the binned data yields virtually
identical results. In the Appendix, we describe our implementation of
Gibbs sampling MAP and how it compares with the much simpler method we
use in this paper.  

Here, we simply bin the data and adopt as our log-likelihood the C-statistic \citep{Cash1979}: 
\begin{eqnarray}
-2{\rm ln} {\cal L}(p|d) = -2 \sum_{i=1}^{N_{bin}}n_i {\rm ln} e_i- e_i -
{\rm ln} n_i !
\label{eqn:cashc}
\end{eqnarray}
where ${\cal L}(p|d)$ is the likelihood of the parameters $p$ given the data $d$, $N_{bin}$ is the number of bins, $n_i$ and $e_i$ are the
observed and expected number of counts in the i-th bin; $n_i,e_i$ are obviously functions of the data $d$ and the unknown model parameters $p$ respectively. It assumes that the number of data points in each bins is Poisson distributed (indeed, it is simply the log of the Poisson likelihood). As we describe in the Appendix, maximizing this statistic produces identical results to more rigorous mixture modeling techniques for large number of data points, when the bin size is sufficiently small. Naively, for a large number of data points one might expect $\chi^{2}$ minimization to work equally well. However, in fitting distributions we are sensitive to the wings of the Gaussian basis functions, when the expected number of counts in a bin is small and the data is therefore Poisson rather than Gaussian distributed. 

With the likelihood specified in Equation \ref{eqn:cashc}, we sample
from the posterior using Metropolis-Hastings MCMC, adapted from CosmoMC \citep{Lewis2002}. Each run draws $\sim 10^5$ samples. The first
30\% are regarded as burn-in and are ignored in the post-analyses. For all
the runs, we visually exam the trace plots to check for convergence. The MCMC analysis yields the
best-fit MAP parameters as well as the full posterior distribution of
parameters, which allows us to estimate confidence intervals. In all
cases, we use non-informative (uniform) priors; the range of possibilities for the turbulent velocity field is
sufficiently large that only very weak priors are justifiable. The only obvious prior we use is $0 < f_{i} < 
1$. Note that there are two identical modes in the likelihood, since
it is invariant under permutation of the mixture indices--the
well-known identifiability or ``label-switching'' problem. Generally, in
a $k$ component mixture, there are $k!$ identical modes in the
likelihood. During the course of a Monte-Carlo simulation, instead of
singling out a single mode of the posterior, the simulation may visit
portions of multiple modes, resulting in a sample mean which in fact
lies in a very low probability region, as well as an unrealistic
probability distribution. We enforce identifiability in a very simple
manner by demanding $\mu_{1} < \mu_{2}$ and hence $s \equiv
\mu_{2}-\mu_{1} >0$. While this is known to sometimes be problematic
\citep{celeux00,jasra05}, in practice it suffices for our simple
models. 

For a large number of data points, the distribution of model parameters becomes asymptotically Gaussian, in which case the Fisher matrix can be used to quickly estimate joint parameter uncertainties
(e.g., \citet{Tegmark1997}).
As a consistency check, we therefore also calculate the Fisher matrix whenever the input model is simple enough to
be expressed analytically. It is defined as: 
\begin{eqnarray}
F_{ij}=-\left<\frac{\partial^{2}\ln{\cal{L}}}{\partial p_i \partial p_j}\right>,
\label{eqn:fisher}
\end{eqnarray}
where $p_i$ is the i-th model parameter. The best attainable
covariance matrix is simply the inverse of the Fisher matrix,
\begin{eqnarray}
C_{ij}=(F^{-1})_{ij},
\end{eqnarray}
and the marginalized error on an individual parameter $p_i$ is
$\sqrt{(F^{-1})_{ii}}$. Differences
between the MCMC and the Fisher matrix error bars generally indicate the
non-Gaussianity of the likelihood surface (or equivalently, that the log-likelihood cannot be truncated at second order in a Taylor expansion).

\section{Idealized Models}
\label{sec:constraints}

\subsection{Two component Gaussian mixture models: General Results}
\label{subsec:general}

\begin{figure}
\begin{tabular}{c}
\rotatebox{-0}{\resizebox{80mm}{!}{\includegraphics{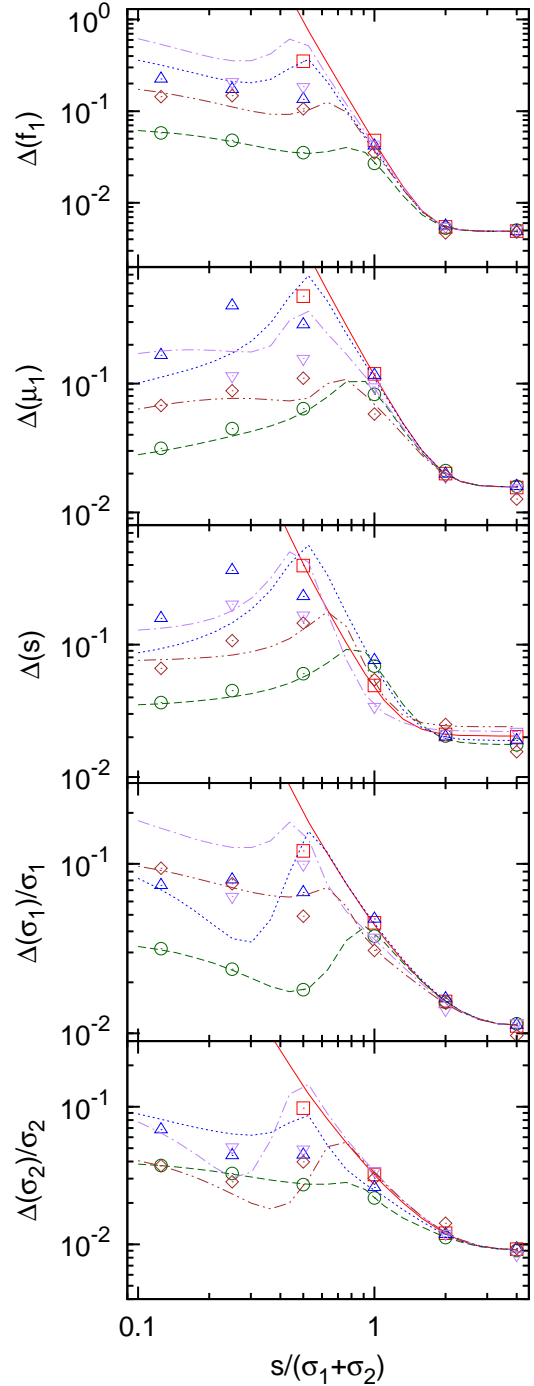}}}
\end{tabular}
\caption{Constraints on the five model parameters with $N_{d}=10^{4}$ data
  points, as a function of $s/(\sigma_{1}+\sigma_{2})$, where $s=\mu_{2}-\mu_{1}$ is the separation between the means. The curves and points
  are the results obtained using the 
  Fisher matrix and MCMC methods, respectively. The 
  dashed lines and circles [green], dotted
  lines and upward triangles [blue], solid
  lines [red], dot-dashed lines and downward
  triangles [purple] and dot-dot-dashed lines and diamonds [brown] correspond
  to $\sigma_{2}=(0.6,0.8,1,1.2,1.4) \sigma_{1}$, respectively.}
\label{fig:separation}
\end{figure}

\begin{figure}
\begin{tabular}{c}
\rotatebox{0}{\resizebox{85mm}{!}{\includegraphics{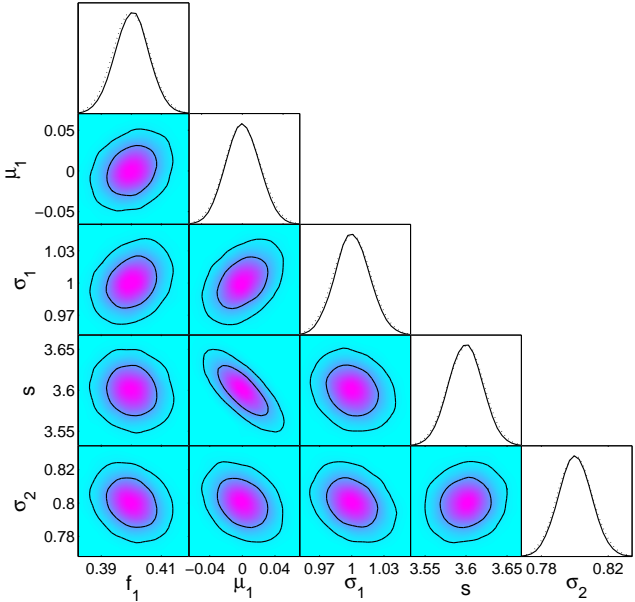}}}
\end{tabular}
\caption{Error contours for a ``SD'' case ($\sigma_1=1$, $\sigma_2=0.8$,
  $s = 2 (\sigma_1+\sigma_2)$): contours depict the
  68\%, 95\% confidence levels for the marginalized distribution; the
  shadings shows the mean likelihood of the samples; the solid and dashed curves
  in the 1-D plots are the fully marginalized posterior and relative
  mean likelihood of the samples, respectively. 
}
\label{fig:corr_sd}
\end{figure}

\begin{figure}
\begin{tabular}{c}
\rotatebox{0}{\resizebox{85mm}{!}{\includegraphics{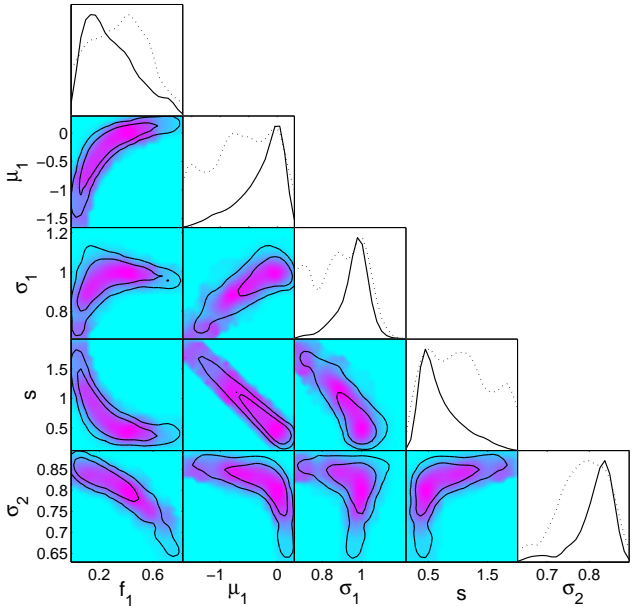}}}
\end{tabular}
\caption{Same as Fig. \ref{fig:corr_sd} but for a ``WD'' case:
  $\sigma_1=1$, $\sigma_2=0.8$, $s=0.25 (\sigma_1+\sigma_2)$.
Note the increased parameter degeneracies.}
\label{fig:corr_wd}
\end{figure}

\begin{figure}
\begin{tabular}{c}
\rotatebox{-90}{\resizebox{60mm}{!}{\includegraphics{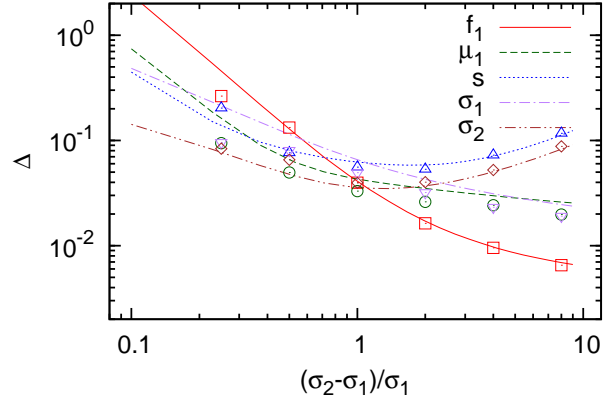}}}
\end{tabular}
\caption{Constraints on the model parameters for a 2 Gaussian mixture as a function of the
  fractional difference in width, when $s=0.2$ and $\sigma_{1}=1$. As in Fig. \ref{fig:separation},
  the lines and points show the results obtained with the Fisher matrix
and MCMC technique, respectively. The solid
  lines and squares [red], dashed lines and circles [green], dotted
  lines and upward triangles [blue], dot-dashed lines and downward
  triangles [purple] and dot-dot-dashed lines and diamonds [brown] are
  the constraints on $f_1$, $\mu_1$, $s$, $\sigma_1$ and $\sigma_2$, respectively}
\label{fig:width}
\end{figure}

\begin{figure}
\begin{tabular}{c}
\rotatebox{0}{\resizebox{80mm}{!}{\includegraphics{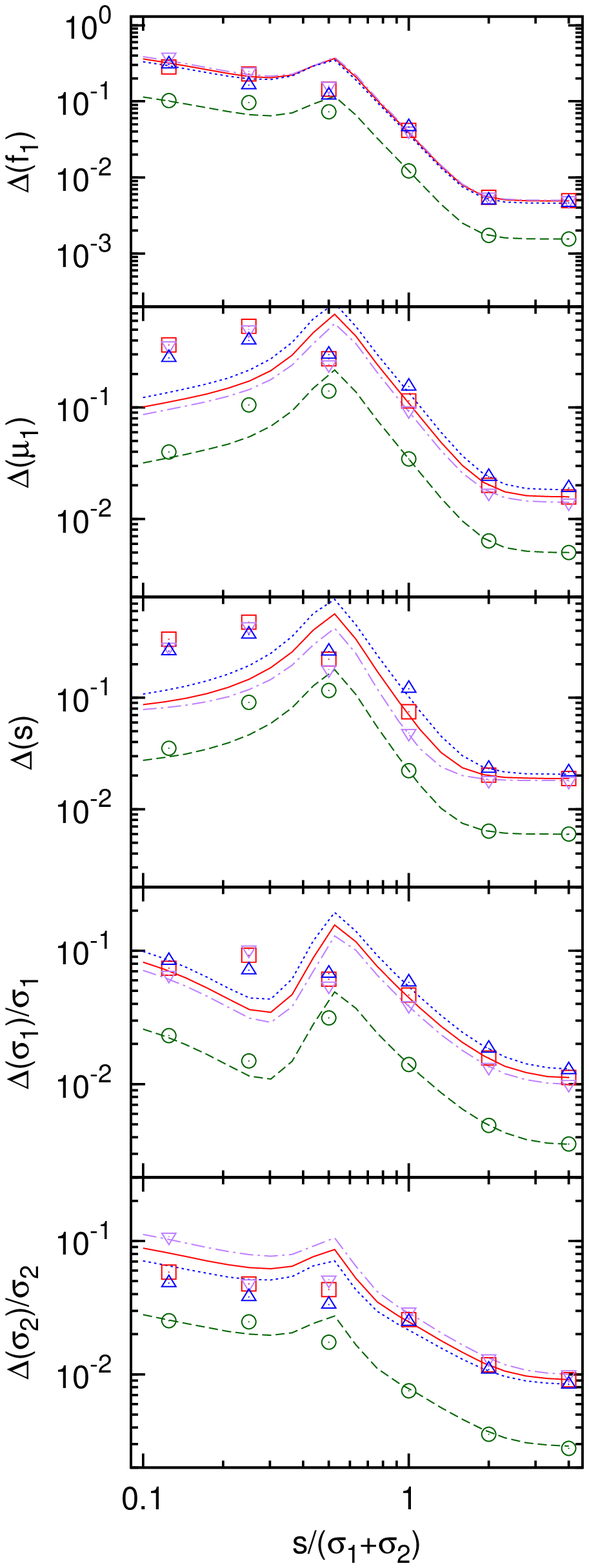}}}
\end{tabular}
\caption{Constraints on the five free parameters for different values
  of $N_d$ and $f_1$. The solid curves and squares are the results for
  the fiducial case: $N_d=10^4$, $\sigma_1=1$, $\sigma_2=1$ and
  $f_1=0.4$; results for $N_d=10^5$, $f_1=0.3$, and $f_1=0.5$ are
  shown with dashed curves and circles, dotted curves and downward
  triangles, and dot-dashed curves and upward triangles, respectively.
  }
\label{fig:frnp}
\end{figure}

Before focusing on the specific application to galaxy clusters, we first consider a
more general problem: how well two Gaussian profiles can be
separated. As mentioned previously, Astro-H data quality is generally
only sufficient to allow solving for the two most dominant mixtures. A
two mixture component is likely the most common scenario, with the most straightforward physical
interpretation. These
results serve to guide and motivate our later discussions. 

Consider therefore the profile:
\begin{eqnarray}
p(x)=\sum_{i=1,2}f_i G(x-\mu_i,\sigma_i),
\label{eqn:profile}
\end{eqnarray}
where $f_i$ is the fraction of each component, while $\mu_i$ and
$\sigma_i$ are the mean and standard deviation of i-th Gaussian
function. Given the constraint $\sum f_i =1$, there are only five model
parameters, which we choose to be: $f_1$, $\mu_1$, $s(\equiv
\mu_2 - \mu_1)$, $\sigma_1$ and $\sigma_2$. Note, $\mu_2$ has been
replaced by $s$ (the separation between the two Gaussians) since
the latter, as we will see more clearly later, usually carries clearer
physical meaning. 

The constraints, expressed in term of standard deviations $\Delta$
throughout this paper, are forecasted with both the MCMC and Fisher matrix
methods 
(for the MCMC runs, they correspond to $68\%, 95\%$ confidence
intervals for $\Delta,2 \Delta$ respectively, even if the parameter
distribution is non-Gaussian). For each model we create a Monte-Carlo
realization with $N_{d}$ data points and forecast constraints for this
data set. Motivated by Table \ref{tbl:clusters}, we assume $N_d=10^4$. 
In general, the standard deviation of the model parameters $\Delta p \propto 1/\sqrt{N_{\rm d}}$, though there are some subtleties---see further discussions below. 
The constraints also depend on how much the two components differ; if they are difficult to distinguish, mixture modeling will fail. For Gaussian components, they may
differ in fraction $f_{i}$, mean $\mu_{i}$ or width $\sigma_{i}$. 
Here, we shall mostly focus on a situation when the mixing fractions are
comparable: $f_{1}=0.4,f_{2}=0.6$, and focus on how mixture separation
can be driven by differences in mean (``separation dominated'', or SD),
or width (``width drive'', or WD).  
In practice, we care mostly about the case when the mixing fractions are roughly comparable, since then the different components are of comparable importance in reconstructing the (emission-weighted) velocity field. As a practical matter, it also becomes increasingly difficult to perform mixture modeling when one component dominates (though see \S \ref{subsec:prospects}). 

The results are shown in Fig. \ref{fig:separation} -
\ref{fig:frnp}. Fig. \ref{fig:separation} shows the constraints as a
function of the separation $s=\mu_{2}-\mu_{1}$, normalized by the sum
of the standard deviations: $s/(\sigma_{1}+\sigma_{2})$. Different line
types 
and point types indicate different values of $\sigma_2=
(0.6, 0.8, 1, 1.2, 1.4) \, \sigma_1$. Note that all five parameters scale with
$s/(\sigma_{1}+\sigma_{2})$ in the same way, with fractional errors
which are all roughly comparable. We can identify three distinct
regimes:  
\begin{itemize}
\item{{\bf Separation Driven} For $s/(\sigma_{1}+\sigma_{2}) \gsim 2$,
    the fractional errors converge to an asymptotic constant value,
    independent of $s/(\sigma_{1}+\sigma_{2})$. In this regime, the
    separation is so large that 
different components could be viewed as individually constrained
without mixing from other components. Except for $\Delta(s)$ (which
depends on $\Delta (\mu_{2}) \sim \sigma_2/\sqrt{f_i N_d}$), this
asymptotic convergence is also independent of $\sigma_{2}$ (i.e., the
relative widths of the distributions don't matter when the separation
is large).  
The asymptotic values for
$\Delta(\mu_i)$, $\Delta(\sigma_i)/\sigma_i$ and $\Delta(f_i)$ are
$\sigma_i/\sqrt{f_i N_d}$, $1/\sqrt{2 f_i N_d}$ and
$\sqrt{{f_i(1-f_i)}/{N_d}}$, respectively; given our $N_{d}=10^{4}$,
this corresponds to $\sim 1\%$ accuracy in parameter constraints. } 
\item{{\bf Hybrid} For $0.3 \lsim s/(\sigma_{1}+\sigma_{2}) \lsim 2$,
    the separation is comparable to the sum of widths. The mixing
    between different components become severe and the quality of
    parameter constraints decrease rapidly with decreasing $s$. Since
    constraints are increasing driven by data points in the tails of
    the respective mixtures (which drive distinguishability), the
    effective number of data points $N_{\rm eff} < N_{d}$
    falls. Strong parameter degeneracies also develop.} 
\item{{\bf Width Driven} When $s/(\sigma_{1}+\sigma_{2}) \lsim 0.3$,
    the separation between the distribution becomes negligible, and
    component separation is driven almost entirely by differences in
    width (note how parameter uncertainties blow up at low $s$ when
    $\sigma_{1}=\sigma{2}$). It is driven to an asymptotic value
    determined by the effective number of data points in the tails of
    the mixtures, $N_{\rm eff}$. } 
\end{itemize} 
The results 
obtained with the Fisher matrix (lines) and MCMC (points) agree well
with each other when the mixtures are easily distinguishable (when $s/(\sigma_{1}+\sigma_{2})$ is large or
$\sigma_2/\sigma_1$ is reasonably far away from 1). Otherwise, discrepancies between these two methods are
clear. These discrepancies are caused by the non-Gaussianity of the
likelihood surfaces and the priors we placed in the MCMC runs. 
In this regime, one therefore cannot use the Fisher matrix approximation to the full error distribution.

Fig. \ref{fig:corr_sd} and \ref{fig:corr_wd} show the marginalized likelihood
  distributions and error contours for two example
  runs. Fig. \ref{fig:corr_sd} is in the SD 
regime ($s = 2 (\sigma_1+\sigma_2)$). The likelihood distributions are
very close to Gaussian, explaining the consistency between the Fisher
matrix and MCMC results. The contours allow direct reading of
correlations among parameters. The strongest correlation is between
$\mu_1$ and $s$. As expected, they are negatively correlated, since
$s=\mu_2-\mu_1$ while the $\mu_i$'s are uncorrelated. Fig. \ref{fig:corr_sd} is in the WD 
regime ($s = 0.25 (\sigma_1+\sigma_2)$). The likelihood distributions
now deviate from Gaussians, and the correlations among parameters are
much stronger. These are all consistent with the facts that the
constraints are worse (due to larger parameter degeneracies) and the
Fisher matrix results are no long in agreement with MCMC results (due
to non-Gaussianity of the likelihood surface).

In Fig. \ref{fig:width}, we show how the constraints vary with 
differences in the Gaussian width in the ``width dominated'' regime. The width of the first component is
fixed to $\sigma_{1}=1$, while the separation $s=0.2$ (note that $s/(\sigma_{1}+\sigma_{2}) \sim 0.2$ is typically the minimal value expected in cluster turbulence when there are no bulk flows, and is due solely to error in the mean; see \S\ref{section:physical_significance}). In general, the
constraints improve as the differences in width 
increase, consistent with intuitive expectations. However, the
constraints on $s$ and $\sigma_2$ turn over around $\sigma_2 \sim
2\sigma_1$, beyond which they increase with $\sigma_2$\footnote{The turnover
does not appear in the last panel of Fig. \ref{fig:separation},
because there the y-axis is $\Delta(\sigma_2)/\sigma_2$ rather than $\Delta(\sigma_2)$}. 
This can be understood as follows: the error on these quantities receive contributions from confusion error (which dominates at low $\sigma_{2}$) and scaling with $\sigma_{2}$ (since $\Delta(\mu_{i}) \sim \sigma_{i}/\sqrt{f_{i} N_{d}}$ and $\Delta \sigma_{i} \sim \sigma_{i}/\sqrt{2 f_{i} N_{d}}$; this dominates at high $\sigma_{2}$).  
On the other hand, the error on the mixing fraction $f_{1}$ scales strongly with the difference in widths, 
since it is driven solely by confusion error. However, for other parameters the scaling is significantly weaker. For most cluster scenarios, the width-driven regime gives relative errors of $\Delta p/p \sim 10\%$, which is still small. 

Fig. \ref{fig:frnp} shows how the constraints vary with $N_d$ and
  $f_1$. The fiducial case (solid curves and squares) is computed
  assuming $N_d=10^4$, $\sigma_1=1$, $\sigma_2=0.8$ and $f_1=0.4$,
  exactly the same as the dotted curves and upward triangles in
  Fig. \ref{fig:separation}. As we increase the $N_d$ by a factor of
  10, most constraints are improved by a factor of $\sqrt{10}$,
  consistent with our expectation that $\Delta p_{i}/p_{i}
\propto 1/\sqrt{N_{d}}$. This is
despite the fact that only in the asymptotic SD case are relative
errors quantitatively given by the Poisson limit $\Delta p_{i}/p_{i}
\approx 1/\sqrt{(f_{i} N_{d})}$. This is because when mixtures overlap
and are in the hybrid/WD regimes, results are driven by the
distribution tails, where the effective number of data points is still
$N_{\rm eff} \propto N_{\rm d}$. 
For $s$ and $\mu_1$, however, MCMC results
show better improvements in the WD regime than factors of
$\sqrt{10}$. This might be due to reduced parameter degeneracies from the
larger number of data points. Varying $f_1$ to 0.3 and 0.5 mildly impacts the
results. As the $f_1$ goes closer to 0.5, constraints improve for most
parameters, except for $f_1$ and $\sigma_2$. Constraints on $f_1$ are
almost unchanged while constraints on $\sigma_2$ are degraded, because
fewer data points are available in the second component to 
constrain $\sigma_2$.

Based on Fig. \ref{fig:separation} and \ref{fig:width}, we can already
anticipate the constraints from Astro-H: when there is significant bulk flow and the modes
have a large relative velocity $v_{\rm bulk} > \sigma_{\rm turb}$, parameters can be constrained to $\sim 1\%$ accuracy (SD regime); when
the relative velocity is small but the widths are different by a
reasonable (a few tens of percents) amount, the parameter estimates are accurate at the $\sim 10\%$ level (WD regime). Given the modeling uncertainties in the physical interpretation of these parameter estimates, such accuracy is more than adequate. Next, we will consider two specific
examples of the SD regime and WD regime respectively.  

\subsection{Application to Clusters: the Single Line Scenario}
\label{subsec:case1}

\begin{figure}
\begin{tabular}{c}
\rotatebox{-90}{\resizebox{100mm}{!}{\includegraphics{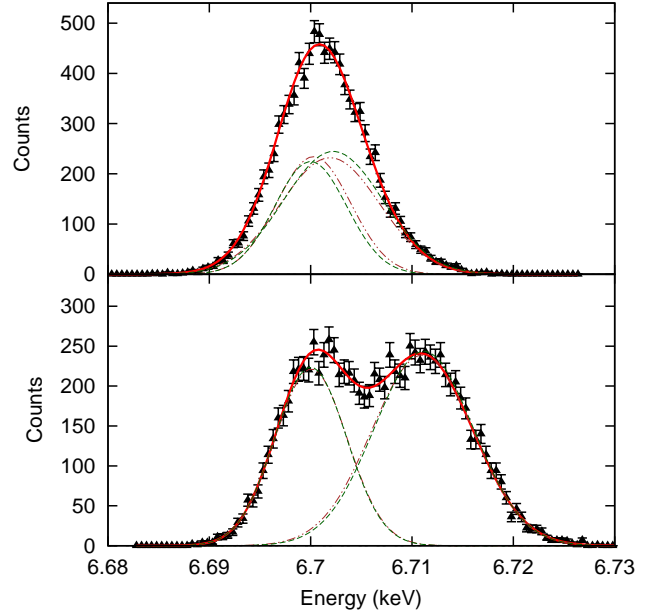}}}
\end{tabular}
\caption{The mock spectra (data points) and best-fit models for the WD
(upper panel) and SD (lower panel) cases in the single line
scenario. The red solid and dot-dashed lines are the input overall spectra and individual components respectively. The green dashed lines are the recovered components. The recovery is remarkably accurate, even when (as in the top panel) the spectra is visually indistinguishable from a single component Gaussian.}
\label{fig:spec_single}
\end{figure}

\begin{figure}
\begin{tabular}{c}
\rotatebox{-90}{\resizebox{100mm}{!}{\includegraphics{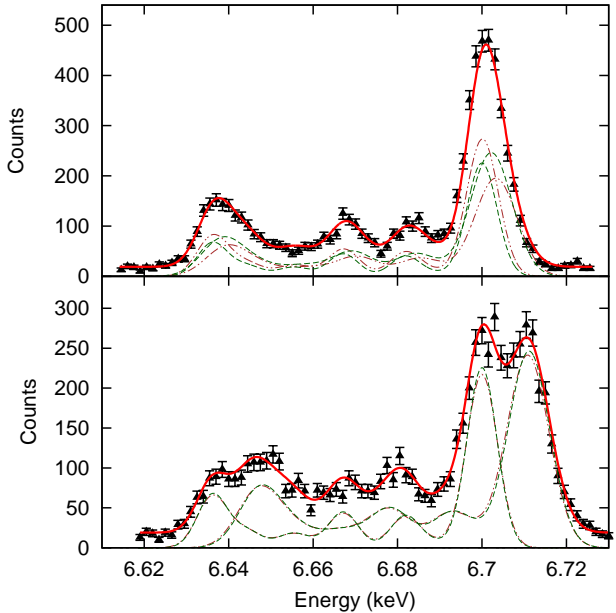}}}
\end{tabular}
\caption{Same as Fig. \ref{fig:spec_single} but for the entire iron line
  complex. The continuum has also been included, assuming a
  metallicity of 0.3 ${\rm Z_{\odot}}$.}
\label{fig:spec_multiple}
\end{figure}

\begin{table*}
  \caption{Input parameters for the WD case and recovered
    best-fit parameters together with their 1-$\sigma$ errors. Also
    shown are the predicted uncertainties using the Fisher matrix technique. Note that $v_{\rm pec},v_{\rm rel}$ are line of sight quantities, while $v_{\rm tb,1}, v_{\rm tb,2}$ are 3D velocity dispersions (assuming $v_{\rm 3D}^{2}=3 v_{\rm 1D}^{2}$).}
    \label{tbl:para_wd}
\begin{center}
\begin{tabular}{c|l| c c c c c}
\hline \hline
&&$f_1$&$v_{pec}$(km/s)&$v_{rel}$(km/s)&$v_{tb,1}$(km/s)&$v_{tb,2}$(km/s)\\
&Input & 0.4 & 0 & 100 & 150 & 300 \\
\hline
\multirow{2}{*}{Single line}&MCMC & $0.46_{-0.12}^{+0.18}$ & $11.34_{-11.32}^{+11.62}$ & $89.48_{-13.85}^{+28.06}$ & $164.89_{-33.97}^{+33.75}$ & $298.82_{-13.34}^{+11.43}$ \\
&Fisher Matrix & (0.12) & (12.99) & (14.06) & (37.03) & (10.75)\\
\hline
\multirow{2}{*}{Multiple lines}&MCMC & $0.42_{-0.09}^{+0.25}$ & $-14.24_{-9.79}^{+17.15}$ & $130.72_{-11.19}^{+65.05}$ & $145.48_{-33.65}^{+42.59}$ & $291.64_{-34.19}^{+5.40}$ \\
&Fisher Matrix & (0.12) & (11.28) & (17.37) & (36.75) & (10.65)\\
\hline
\multirow{1}{*}{Multiple lines}&MCMC & $0.64_{-0.13}^{+0.22}$ & $5.44_{-9.08}^{+18.10}$ & $147.88_{-34.85}^{+124.25}$ & $180.16_{-21.43}^{+26.58}$ & $308.14_{-83.92}^{+13.44}$ \\
plus continuum&Fisher Matrix & (0.19) & (15.36) & (27.04) & (53.07) & (17.19)\\

\hline
\end{tabular}
\end{center}
\end{table*}

\begin{table*}
  \caption{Same as Table \ref{tbl:para_wd} but for the SD case.}
    \label{tbl:para_sd}
\begin{center}
\begin{tabular}{c|l| c c c c c}
\hline \hline
&&$f_1$&$v_{pec}$(km/s)&$v_{rel}$(km/s)&$v_{tb,1}$(km/s)&$v_{tb,2}$(km/s)\\
&Input & 0.4 & 0 & 500 & 150 & 300 \\
\hline
\multirow{2}{*}{Single line}&MCMC & $0.40_{-0.02}^{+0.01}$ & $2.17_{-6.40}^{+7.28}$ & $493.65_{-4.82}^{+4.54}$ & $152.71_{-5.93}^{+7.62}$ & $310.98_{-5.33}^{+6.65}$ \\
&Fisher Matrix & (0.01) & (6.40) & (4.52) & (11.45) & (9.77)\\
\hline
\multirow{2}{*}{Multiple lines}&MCMC & $0.41_{-0.01}^{+0.01}$ & $2.47_{-4.72}^{+6.18}$ & $505.28_{-4.13}^{+3.96}$ & $148.71_{-9.61}^{+14.19}$ & $294.13_{-9.43}^{+8.84}$ \\
&Fisher Matrix & (0.01) & (5.73) & (4.15) & (12.39) & (9.24)\\
\hline
\multirow{1}{*}{Multiple lines}&MCMC & $0.40_{-0.02}^{+0.02}$ & $-0.01_{-8.19}^{+7.88}$ & $486.43_{-4.98}^{+4.90}$ & $167.92_{-16.89}^{+15.33}$ & $309.75_{-14.03}^{+14.04}$ \\
plus continuum&Fisher Matrix & (0.02) & (6.72) & (4.61) & (15.75) & (12.50)\\
\hline
\end{tabular}
\end{center}
\end{table*}

%
%


We begin our discussion of mixture modeling of cluster emission line spectra with the simplest case. For now we ignore
line blending and continuum emission, and only consider one emission
line -- the He-like iron line at 6.7 keV. Again, we assume the PDF is
composed of two Gaussian components. Most of these assumptions will be
relaxed later. We assume the cluster is isothermal with a temperature
of 5 keV. The assumption of an isothermal
distribution is of course somewhat crude for the entire cluster. However,
for nearby clusters, the emission-weighted spectrum is accumulated from a small
area where temperature variations are generally mild ($< 0.5$ keV). 
Moreover,
our results are not very sensitive to the 
temperature distribution. 
We express our results in terms of the bulk peculiar velocity of 
the first component ($v_{pec}$), the relative velocity 
between the two components ($v_{rel}$), and the 3D turbulent velocity dispersions of each component
($v_{\rm tb,1}$ and $v_{\rm tb,2}$). We assume isotropic turbulence,
so the line of sight velocity dispersion is $v_{\rm
  tb,i}/\sqrt{3}$. They are related to the Gaussian PDF via:
\begin{eqnarray}
\mu_1&=&\nu_0+\nu_0\frac{v_{pec}}{c},\\\nonumber
s&=&\nu_0\frac{v_{rel}}{c},\\\nonumber
\sigma_i&=&\sqrt{\sigma_{tb,i}^2+\sigma_{ther}^2+\sigma_{instr}^2},\\\nonumber
\sigma_{tb,i}&=&\nu_0\frac{v_{tb,i}}{\sqrt{3}c},\\\nonumber
\sigma_{ther}&=&\frac{\nu_0}{c}\sqrt{\frac{kT}{Am_p}},\\\nonumber
\label{eqn:profile}
\end{eqnarray}
where $\nu_0$ is the line frequency in the rest frame,
$\sigma_{instr}$ is the standard deviation of instrumental noise
(FWHM/2.35), $A$ is the atomic weight of iron and $m_p$ is the proton mass. 
In our WD 
example, we assume $(v_{\rm tb,1},v_{\rm tb,2})=(150, 300) \, {\rm km
  \, s^{-1}}$, and $v_{\rm rel}=100 \, {\rm km \, s^{-1}}$. For the SD
example, we assume the same $v_{\rm tb,1},v_{\rm tb,2}$, but $v_{\rm
  rel}=500 \, {\rm km \, s^{-1}}$. In all cases, we assume the bulk
velocity zero-point $v_{\rm pec}=0 \, {\rm km \,
  s^{-1}}$.\footnote{Note that if the redshift of the collisionless
  component of the cluster (which does not participate in gas bulk
  motions) can be determined to high accuracy by spectroscopy of
  numerous galaxies, then $v_{\rm pec}, v_{\rm rel}-v_{\rm pec}$ give the
  line of sight bulk velocities with respect to the cluster potential
  well. For instance, for nearby clusters where $N_{\rm
    gal} \sim 400$ galaxy redshifts have been measured,  
the relative error in the center of mass redshift is $\sim 1000 {\rm km \, s^{-1}}/(\sqrt{3}\sqrt{N_{\rm gal}}) \sim 30 {\rm km \, s^{-1}}$. 
Otherwise, only $v_{\rm rel}$ (the relative bulk velocity) is of physical significance.}
Sloshing in the cluster potential well generally results bulk motions
with transonic Mach numbers \citep{markevitch07}, so such a value is
realistic for a 5 keV cluster (with sound speed $c_{\rm s} \sim 1000
\, {\rm km \, s^{-1}}$) along an arbitrary line of sight---indeed,
such velocities are found in the simulated cluster in
\S\ref{subsec:example}.  
With these assumptions, the widths of the first
and second component, including instrumental, thermal and turbulent broadening, are 3.54 and 4.87 eV; the offsets
between peaks are 1.12 and 11.17 eV for the WD and SD cases, respectively. These parameter choices correspond to $s/(\sigma_{1}+\sigma_{2})=(0.13,1.3)$ respectively, and thus can be compared to expectations from Fig. \ref{fig:separation}. Note that the SD case is not quite in the asymptotic regime $s/(\sigma_{1}+\sigma_{2}) \gsim 3$ yet (where the relative errors would be $\sim 1/\sqrt{f_{i} N_{d}} \sim 1\%$), but it is fairly close. 

The mock spectra and best-fit models for $10^4$ photons are shown in
Fig. \ref{fig:spec_single}. The best-fit parameters and their
uncertainties are listed in the first row of Table \ref{tbl:para_wd} and
\ref{tbl:para_sd}. In accordance with expectations from \S\ref{subsec:general}, component recovery is remarkably accurate. Even is the WD case, which is visually indistinguishable from a single Gaussian (see top panel of Fig. \ref{fig:spec_single}), the decomposition into the original mixtures is very good, and most velocities are constrained to within $\sim 10-30 \, {\rm km \, s^{-1}}$, which is significantly higher accuracy than needed to model the physical effects of bulk motions and turbulence in the cluster. This showcases the great
potential of high spectral resolution instruments. Of particular interest is the constraint on the mixing fraction, which
is a very good indicator of our ability to separate different
components. A confident detection of multiple components should have
$f_1/\Delta(f_1)$ larger than a few, i.e., the best-fit fraction should be at
least a few $\sigma$ away from ``non-detection'' ($f_1$ or $f_2$ equal
to 0). In the single line scenario, the $1-\sigma$ error of $f_1$ is 0.01 and 
0.12 in the SD and WD cases, respectively, consistent with our
expectations from discussions in the previous section. However, a large 
fraction of the constraints in the WD case is from the tails, and could 
easily be affected by continuum emission (see discussion
below). Also note the general consistency between Fisher matrix and MCMC
techniques, indicating the Gaussian shape of the likelihood
surface for this scenario. 

\subsection{The Impact of Multiple Lines and Continuum}
\label{subsec:case2}

In this section, we consider the impact of multiple lines and
continuum emission. Iron lines appear as a line complex between 6.6
and 6.75 keV, and these lines inevitably blend together. Multiple
lines have two competing effects. First, taking all lines into account--all of which have identical mixture decompositions--means more photons, which reduces shot noise in parameter estimates. The photons from the entire line complex is about twice that from the He-like iron line alone. Secondly, as different
lines blend together, information contained in the shape of individual
lines is partly lost due to blending in the line wings. The latter are crucial to driving parameter estimation in the hybrid and WD cases (note however, from Fig. \ref{fig:spec_multiple} that the lowest and highest energy lines in the complex have low/high energy line wings respectively which are unaffected by blending. This is particularly important in the case of the high energy He-like line, which is by far the strongest line in the complex). These two factors have opposite effects on the
constraints. As in the previous sub-section, 
we run MCMC chains and Fisher matrices to estimate the constraints.
The properties of the line complex were taken from ATOMDB database\footnote{http://www.atomdb.org/}
(v. 2.0.1). To save computing time, we only included the ten strongest lines
lines. Fisher matrix estimates including more lines show negligible
difference. 

The results are listed in the second row of Table \ref{tbl:para_wd} and
\ref{tbl:para_sd}. In the SD case, the constraints estimated using both
MCMC and Fisher matrix techniques are very close to those in the
single line scenario, indicating almost total cancellation between
the effects just mentioned. In the WD case, the constraints from the
Fisher matrix technique are again close to the single line
scenario. However, the results from MCMC runs show asymmetry, and in
general, the constraints are worse than in the single line scenario. Line
blending seems to make the likelihood surface significantly non-Gaussian.

Next, we include the effect of continuum emission. Continuum acts as a source of background noise. Even though we can measure and subtract the continuum, doing so introduces shot noise, particularly in the line wings when Fe line emission and continuum brightness can become comparable, or continuum emission could even dominate. The relative level of
continuum and Fe line emission is controlled by metallicity; larger metallicities imply brighter lines. The mean
metallicity of clusters is typically ${\rm Z} \sim 0.3 \, {\rm Z_{\odot}}$, which we shall assume, though the metallicity in the cluster center is often higher due to contributions from the cD galaxy. We apply our mixture model incorporating both the effects of line blending and continuum; the results are shown in Table \ref{tbl:para_wd} and
\ref{tbl:para_sd}, and in Fig. \ref{fig:spec_multiple} (for
the purpose of clarity, only 1/3 of the data points are shown in this figure). 
The results are as one might expect. In the SD case, the constraints are only slightly worsened, since the mixtures 
are clearly separated, and almost all the $\sim f_{i} N_{d}$ points in a given mixture can be used for parameter estimates; only a small fraction in the line tails are contaminated by line blending and the continuum. The constraints in the WD case are more badly affected, since the constraints in this case are largely drawn from the tails; here, the
differences between the MCMC and Fisher matrix techniques are also further
enlarged.  
The presence of the continuum and line blending limit the
domain of the WD regime, which is no longer strictly independent of
$s/(\sigma_{1}+\sigma_{2})$. For instance, if we assume $v_{\rm rel} =
50 \, {\rm km \, s^{-1}}$ (corresponding to
$s/(\sigma_{1}+\sigma_{2})= 0.067$), the MCMC simulations fail to
converge). They thus limit our ability to constrain components with
small separations, though in practice such small separations should be
rare. 


\subsection{Model Selection: When is a Mixture Model fit Justified?}
\label{subsec:prospects}

\begin{figure}
\begin{tabular}{c}
\rotatebox{-90}{\resizebox{60mm}{!}{\includegraphics{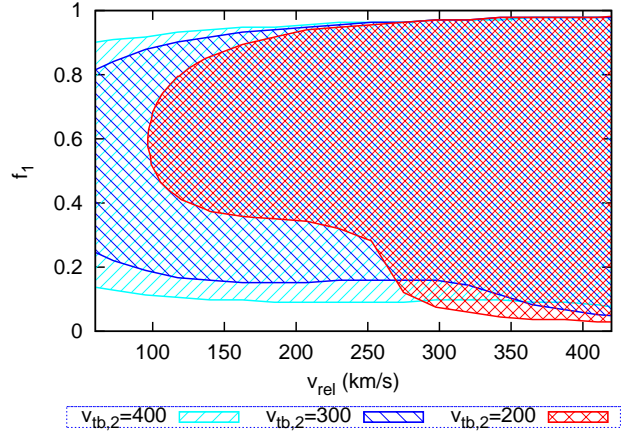}}}
\end{tabular}
\caption{Model selection: regions of the $f_1$ - $v_{rel}$ plane where the double component
  model is preferred according to the BIC, for $v_{\rm tb,1}=100 \, {\rm km \, s^{-1}}$, $v_{\rm tb,2}=(200,300,400) \, {\rm km \, s^{-1}}$, and $f_{1}=0.4$, $v_{\rm pec}=0 \, {\rm km \, s^{-1}}$. }
\label{fig:select1}
\end{figure}
 
\begin{figure}
\begin{tabular}{c}
\rotatebox{-90}{\resizebox{60mm}{!}{\includegraphics{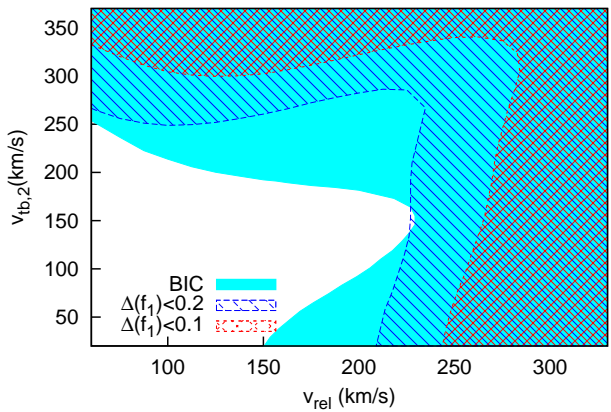}}}
\end{tabular}
\caption{Model selection: shaded regions shows the regions of $(v_{\rm rel},v_{\rm tb,2})$ parameter space where the double
  component model is preferred according to BIC, while the hatched
  regions are where mixing fraction is accurately constrained: $\Delta(f_{1}) < (0.2,0.1)$ (cyan and purple hatches respectively). All other parameters are as in Fig. \ref{fig:select1}.}
\label{fig:select2}
\end{figure}

Thus far, we have only considered how accurately mixture model parameters can be constrained. However, this begs the question  of whether a mixture model approach is justified at all, particularly when (as in the WD case) the observed emission line is visually indistinguishable from a single Gaussian. Introducing additional parameters will always result in an improved fit, even when these parameters are largely irrelevant and of little physical significance. This is essentially a model selection problem. We use information criteria (e.g., see \citet{liddle04}) which penalize models with more parameters, to identify preferred models. While they have solid underpinnings in statistical theory, fortunately, they have very simple analytic expressions. In this paper, we use the Bayesian Information Criterion (BIC; \citet{Schwarz1978}): 
\begin{eqnarray}
BIC\equiv -2 \ln{\cal L}_\mathrm{max}+k\ln N
\label{eqn:bic}
\end{eqnarray}
where ${\cal L}_\mathrm{max}$ is the maximum likelihood achievable by
the model, $k$ is the number of free parameters, and $N$ is the number
of data points; the preferred model is one which minimizes BIC. The BIC comes from the Bayes factor \citep{jeffreys61}, which gives the posterior odds of one model against another. We use it over the closely related Akaike Information Criterion (AIC; \citet{akaike74}), which places a lower penalty on additional model parameters. Thus, we adopt a conservative criterion for preferring mixture models. The absolute value of the BIC has no significance, only the relative value between models. A difference of 2 is regarded as positive evidence, and of 6 or more as strong evidence, to prefer the model with lower BIC \citep{jeffreys61,mukherjee98}. Note that the BIC does not incorporate prior information. This is possible with the more sophisticated notion of Bayesian evidence (e.g., \citet{mackay03}), but involves expensive integrals over likelihood space, and is unnecessary in our case since we adopt uninformative priors.

We aim to distinguish the double component model with $k=5$ (free
parameters: $(v_{\rm pec}, v_{\rm rel}, f_{1}, v_{\rm tb,1}, v_{\rm
  tb,2}$), against the single component model with $k=2$ (free
parameters: $\mu, \sigma$). We create simulated data sets which have
two underlying components, and see which regions of parameter space
the BIC will correctly prefer the two component model.  
Our simulated line profiles incorporate the additional effects of
thermal and instrumental broadening, continuum, and line blending. 
Rather than exploring the full 5 dimensional space, we explore the most interesting subspace to see where model selection is effective. In Fig. \ref{fig:select1} we explore model selection in the $f_1$ - $v_{\rm rel}$ plane, for $v_{\rm tb,2}=(200,300,400) \, {\rm km \, s^{-1}}$, and $v_{\rm pec}=0 \, {\rm km \, s^{-1}}$, $f_{1}=0.4$, $v_{\rm tb,1}=100 \, {\rm km \, s^{-1}}$. The plot shows where the BIC for the double component fit is smaller than that for the single component fit (note that the BIC is obtained by allowing for variation in all fitted parameters; we are just plotting model selection in a subspace). 
When $v_{\rm tb,2}=400 \, {\rm km \, s^{-1}}$, all values of $v_{\rm rel}$ and all $0.1 < f_{1} < 0.9$ permit correct selection of the double component model. The result is very similar for $v_{\rm tb,2}=300 \, {\rm km \, s^{-1}}$, but for $v_{\rm tb,2}=200 \, {\rm km \, s^{-1}}$, if both $f_{1}, v_{\rm rel}$ assume low values, the double component model is not preferred. Overall, it is reassuring to see that model selection is not very sensitive to $f_{1}$, since we previously restricted our studies to $f_{1}=0.4$. Thus, even if a smaller fraction of the emission weighted volume has a markedly different velocity structure, it will be detectable in the spectrum. In Fig. \ref{fig:select2}, we show the regions of $(v_{\rm rel},v_{\rm tb,2})$ parameter space where the double
  component model is preferred according to BIC. Overall, as expected,
  the mixtures can be distinguished if $v_{\rm rel}$ or $v_{\rm tb,2}$
  are large; for Astro-H and with the adopted parameters, this is of
  order $200 \, {\rm km \, s^{-1}}$. In addition, we show the regions
  where the mixing fraction $f_{1}$ is accurately constrained to
  $\Delta (f_{1}) < (0.1,0.2)$, since the error on the mixing fraction
  should be a good indicator of our ability to distinguish
  mixtures. We use the Fisher matrix formalism to calculate these
  constraints. The results are qualitatively similar that obtained
  with the BIC, though somewhat more restrictive.

\subsection{Non-Gaussian Mixture Components}
\label{subsec: nongaussianity}

\begin{figure}
\begin{tabular}{c}
\rotatebox{-90}{\resizebox{100mm}{!}{\includegraphics{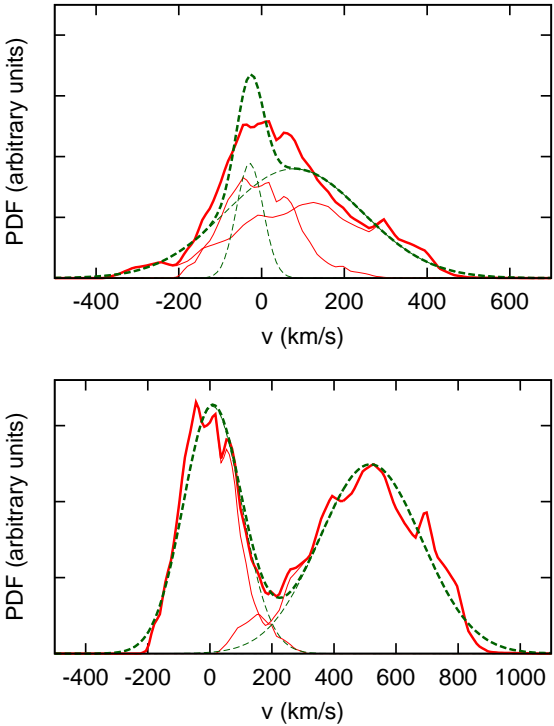}}}
\end{tabular}
\caption{The velocity PDFs in the WD (upper panel) and SD (lower
  panel) cases. The solid (red) and dashed (green) curves are the input and
  recovered PDFs, respectively. The thick curves are the overall PDFs,
while the thin curves show the individual components. }
\label{fig:nongaus}
\end{figure}

\begin{table*}
  \caption{Non-Gaussian mixture components: input parameters (obtained by shifting and rescaling the non-Gaussian mixtures) and recovered
    best-fit parameters together with their 1-$\sigma$ errors, for both the WD and SD cases, as in Fig. \ref{fig:nongaus}. 
Note that the Fisher matrix results--which require an analytic likelihood--assume Gaussian mixtures, and hence are the same as Tables \ref{tbl:para_wd} and \ref{tbl:para_sd}.}
    \label{tbl:nongaus}
\begin{center}
\begin{tabular}{c|l| c c c c c}
\hline \hline
&&$f_1$&$v_{pec}$(km/s)&$v_{rel}$(km/s)&$v_{tb,1}$(km/s)&$v_{tb,2}$(km/s)\\
\hline
\multirow{3}{*}{WD}&Input & 0.4 & 0 & 100 & 150 & 300 \\
&MCMC & $0.17_{-0.01}^{+0.49}$ & $-33.18_{-11.54}^{+37.53}$ & $112.49_{-10.96}^{+89.48}$ & $63.80_{-14.55}^{+138.51}$ & $284.99_{-39.82}^{+4.86}$ \\
&Fisher Matrix & (0.19) & (15.36) & (27.04) & (53.07) & (17.19)\\
\hline
\multirow{3}{*}{SD}&Input & 0.4 & 0 & 500 & 150 & 300 \\
&MCMC & $0.43_{-0.02}^{+0.01}$ & $8.37_{-7.44}^{+5.83}$ & $509.22_{-4.49}^{+4.49}$ & $162.28_{-17.94}^{+12.09}$ & $284.55_{-10.77}^{+13.81}$ \\
&Fisher Matrix & (0.02) & (6.72) & (4.61) & (15.75) & (12.50)\\
\hline
\end{tabular}
\end{center}
\end{table*}

All the preceding discussions are based on the assumption that the
PDFs of individual components are Gaussian, which is not true in general. As we see in
Fig. \ref{fig:spectrum}, individual mixtures show deviations from Gaussianity, i.e. Gaussians are a good but imperfect set of basis functions. In principle, this can be dealt with by fitting higher order mixture models, but in practice the data quality from Astro-H does not allow this; parameter estimation becomes unstable and large degeneracies develop, particularly since the higher order mixtures generally have low mixing fractions $f_{i}$. Unless their velocity means or widths are very different, the physical interpretation of these additional components is also more difficult. Here we construct a simple toy model to isolate the effects of non-Gaussian components. As there are many flavors of non-Gaussianity, the results we show are meant to be illustrative rather than definitive.

To this end, we extract PDFs from a simulated
relaxed cluster, use them as the ``basis'' PDFs of individual components, resize
and combine them to produce a composite PDF, which is in turn used to
generate mock spectra. The ``basis'' PDFs are extracted from a simulation
by \citet{Vazza2010}, which the authors kindly made public; 
we sample different PDFs by looking along different lines of sight.  
The cluster, labeled as E14, has a mass of ${\rm M} \sim
10^{15}~\msun$ and experienced its latest major merger at
$z>1$. Due to shot noise which arises from the finite resolution (25
${\rm kpc} \, h^{-1}$) of this simulation--which results in a small
number of cells--we are 
forced to extract the emission weighted velocity PDFs from a large volume of
$400\times 400 \times 1000~{\rm kpc}^3$. 
The PDFs are
shifted (to match means), linearly rescaled (to match variances) and combined to produce the same WD and SD cases in
\S~\ref{subsec:case2}.\footnote{We emphasize that this procedure is {\it not} meant to simulate what an realistic observation would see, which we treat in \S\ref{sec:application}. It is a toy model in the spirit of the preceding sub-sections, where we use simulations to generate non-Gaussian mixture components.} 
We then convolve the composite PDFs with thermal
broadening and instrumental noise for the entire Fe line complex, and
add continuum 
to produce mock spectra. Finally, we
fit the mock spectra 
to separate and constrain the two components. The results are shown in 
Fig. \ref{fig:nongaus} and Table \ref{tbl:nongaus}. In
Fig. \ref{fig:nongaus}, the solid (red) curves are the input PDFs
while the dashed (green) curves are the best-fit model. The thick and
thin curves are the total PDF and individual components,
respectively 
(note that because we display the velocity PDF rather than the
spectrum, the multiple lines in the Fe complex, as well as the
continuum and thermal/instrumental broadening, are not shown. However,
all these effects are included in the simulations). In the SD case
(lower panel), the two 
components are recovered almost perfectly. In the WD case (upper
panel), however, there are some discrepancies between the input and
output PDFs. The same conclusion can be drawn from Table
\ref{tbl:nongaus}; in the WD case, the best-fit values of $f_1$
and $v_{tb,1}$ are somewhat different from the input values. However,
they are still within the (large) errors. Comparing Table \ref{tbl:nongaus} with Table
\ref{tbl:para_wd} and \ref{tbl:para_sd}, we see that at least in this case, 
non-Gaussian components have limited effect on the results. Note the
strong discrepancy between MCMC and Fisher matrix error bars in both
cases, and in particular the strong asymmetry in MCMC errors. 
We repeated the same exercise several times with PDFs randomly drawn 
along different lines of sight
from the same
simulation. In most attempts, we are able to recover the input
parameter values within the uncertainties. Thus, conclusions
based on Gaussian components are still applicable when the true
PDFs deviate from Gaussianity by a reasonable amount. 
Instrumental and thermal broadening, which {\it gaussian}, effectively
smooth out small scale deviations from Gaussianity.

\section{Results from Numerical Simulations}
\label{sec:application}

\subsection{Cold Front Cluster}

\begin{figure}
\begin{tabular}{c}
\rotatebox{-90}{\resizebox{100mm}{!}{\includegraphics{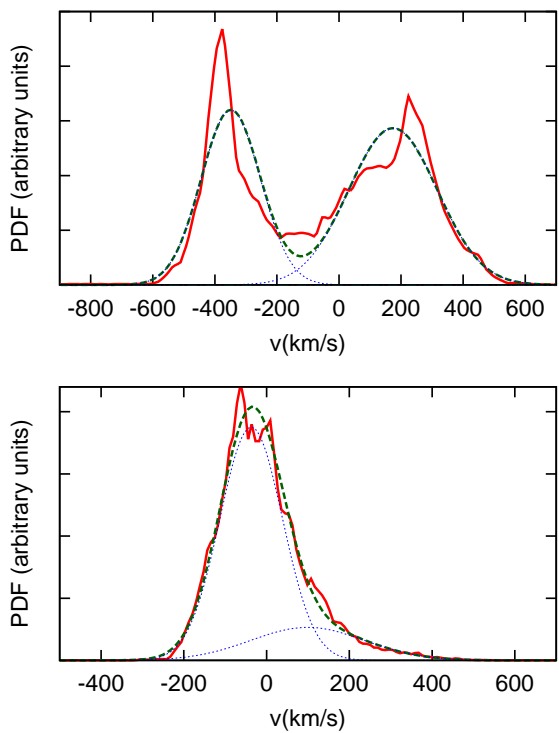}}}
\end{tabular}
\caption{``Cold front'' cluster: the solid (red) curves are the same velocity PDFs as in
  Fig. \ref{fig:spectrum}. The dashed (green) curves are the recovered
PDFs from the best-fit models, and the dotted (blue) curves are the
individual components. Numerical values of the fit parameters are in Table \ref{tbl:app1}.}
\label{fig:app1}
\end{figure}

\begin{table*}
  \caption{``Cold front'' cluster: best-fit parameters and their uncertainties for the
    PDFs in Fig. \ref{fig:app1}, obtained using the Enzo simulation described in \S\ref{subsec:example}. Case 1 and 2 are the top and bottom panels of Fig. \ref{fig:app1} respectively. The ``true values'' are
    obtained by fitting the PDF directly,
    while the recovered values are obtained from the mock spectrum of
    $10^4$ photons, which includes line blending, thermal and instrumental broadening,
   and continuum emission.}
    \label{tbl:app1}
\begin{center}
\begin{tabular}{c | l | c c c c c}
\hline \hline
&&$f_1$&$v_{pec}$(km/s)&$v_{rel}$(km/s)&$v_{tb,1}$(km/s)&$v_{tb,2}$(km/s)\\
\hline
\multirow{2}{*}{Case 1}&True values & $0.40_{-0.01}^{+0.01}$ & $-359.53_{-1.93}^{+1.73}$ & $517.15_{-2.76}^{+2.24}$ & $155.00_{-2.68}^{+2.49}$ & $269.39_{-3.02}^{+4.04}$ \\
&Recovered& $0.43_{-0.01}^{+0.01}$ & $-348.67_{-6.00}^{+6.34}$ & $522.79_{-4.82}^{+3.31}$ & $165.52_{-16.27}^{+13.16}$ & $247.04_{-9.66}^{+12.70}$ \\
\hline
\multirow{2}{*}{Case 2}&True values& $0.78_{-0.02}^{+0.02}$ & $-40.49_{-1.52}^{+1.70}$ & $142.10_{-6.40}^{+11.28}$ & $128.32_{-2.01}^{+2.17}$ & $220.24_{-6.93}^{+4.22}$ \\
&Recovered& $0.80_{-0.26}^{+0.11}$ & $-37.14_{-13.42}^{+9.09}$ & $136.29_{-60.12}^{+78.13}$ & $131.28_{-38.74}^{+11.02}$ & $237.73_{-53.31}^{+23.30}$ \\
\hline
\end{tabular}
\end{center}
\end{table*}

Finally, we apply our tool to cluster simulations. In the first
example, we attempt to recover the velocity PDFs shown in
Fig. \ref{fig:spectrum}, which derives from a cosmological ENZO
simulation of a cold front cluster. These two cases, which come from
different lines of sight through the same cluster, correspond to
$s/(\sigma_{1}+\sigma_{2}) = {1.42}$ and
$s/(\sigma_{1}+\sigma_{2}) = {0.42}$ 
respectively, i.e. in the ``separation-driven'' and ``width-driven''
regimes. We first fit the PDFs with a mixture model when no sources of
noise or confusion are present, to derive the ``true'' parameter
values.  
We then generate mock spectra by adding thermal and instrumental broadening,
continuum emission and line blending to the PDFs,
and 
then apply mixture modeling to the results. The results are given in Fig. \ref{fig:app1} and Table \ref{tbl:app1}.
Overall, the results are very good. The best-fit models
successfully recover the general features of the PDFs, and accurate
parameter estimates with uncertainties which are consistent with our
estimates from the
toy models -- on the order of $\sim 10\%$ for the width-drive case
(case 2) and $\sim 1\%$ for the separation-driven case (case 1). No
systematic biases appear to be present. As we discussed in
\S\ref{section:physical_significance}, these parameters all have
physical significance: $v_{\rm tb,i}$ relates to the turbulent energy
density in each component, $f_{i}$ to the emission weighted volume
fraction of each component, and $v_{\rm rel}$ to the bulk velocity
shear between them. We also applied the single component model to the
same mock spectra, and compared the BIC values. In both cases, the double
component model is preferred (case 1: $BIC_{\rm double}-BIC_{\rm
  single}= -1002$; case 2: $BIC_{\rm double}-BIC_{\rm single}= -10$).


\subsection{AGN Feedback Cluster} 

\begin{figure}
\begin{tabular}{c}
\rotatebox{0}{\resizebox{85mm}{!}{\includegraphics{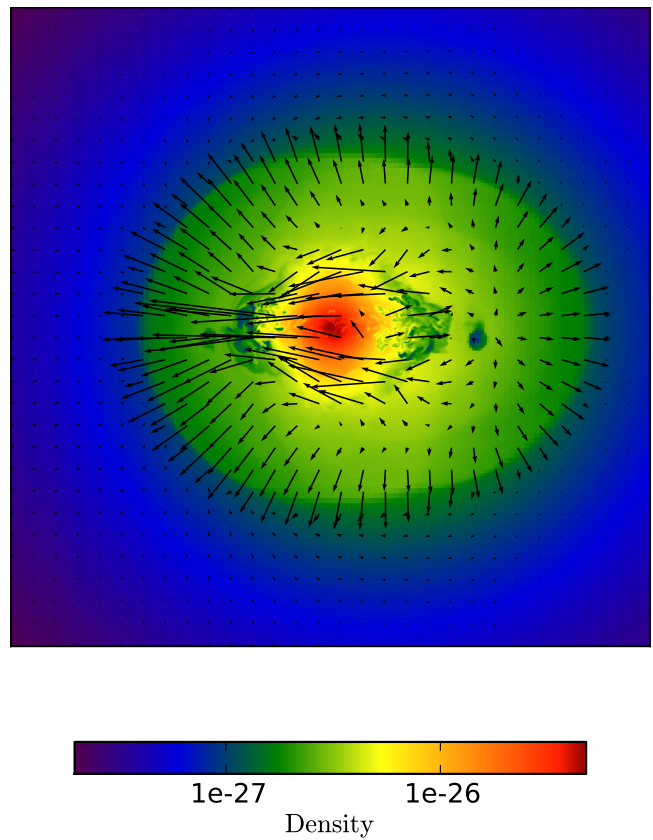}}}
\end{tabular}
\caption{Density map and velocity field on the $y-z$ plane. The size
  of the figure is 1 Mpc; the bulk motion along the y-direction has
  been substracted out. 
}
\label{fig:agnmap}
\end{figure}

\begin{figure}
\begin{tabular}{c}
\rotatebox{-90}{\resizebox{60mm}{!}{\includegraphics{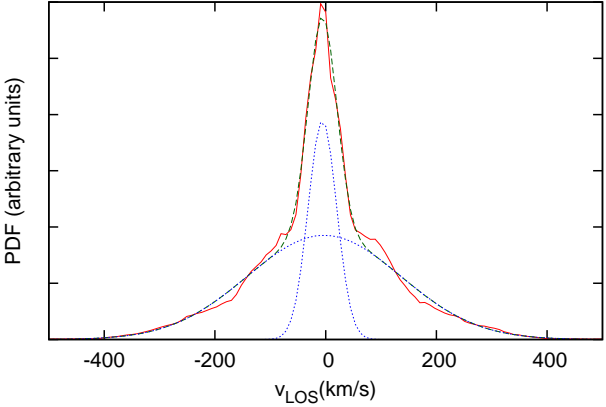}}}
\end{tabular}
\caption{``AGN feedback'' cluster: the solid (red) curves are the velocity PDFs from the simulation. The dashed (green) curves are the recovered
PDFs from the best-fit models, and the dotted (blue) curves are the
individual components. Numerical values of fits are in Table \ref{tbl:app2}.}
\label{fig:agn}
\end{figure}

\begin{figure}
\begin{tabular}{c}
\rotatebox{0}{\resizebox{85mm}{!}{\includegraphics{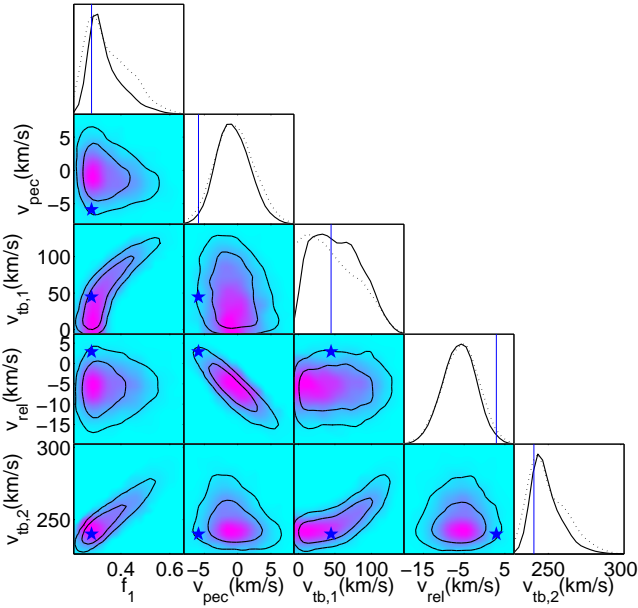}}}
\end{tabular}
\caption{Error contours for the ``AGN feedback'' case: contours depict the
  68\%, 95\% confidence levels for the marginalized distribution; the
  shadings shows the mean likelihood of the samples; the solid and dashed curves
  in the 1-D plots are the fully marginalized posterior and relative
  mean likelihood of the samples, 
respectively. The stars and vertical lines label the
  positions of the true values.} 
\label{fig:agn_2D}
\end{figure}

\begin{table*}
  \caption{``AGN feedback'' cluster: best-fit parameters and their uncertainties for the
    simulated PDF in Fig. \ref{fig:agn}. The ``true values'' are
    obtained by fitting the PDF directly,
    while the recovered values are obtained from the mock spectrum of
    $10^5$ photons, which includes line blending, thermal broadening,
    instrument noise, and continuum emission.}
    \label{tbl:app2}
\begin{center}
\begin{tabular}{ l | c c c c c}
\hline \hline
&$f_1$&$v_{pec}$(km/s)&$v_{rel}$(km/s)&$v_{tb,1}$(km/s)&$v_{tb,2}$(km/s)\\
\hline
True values & $0.28_{-0.01}^{+0.01}$ & $-5.89_{-0.84}^{+0.78}$ & $2.91_{-1.72}^{+2.16}$ & $44.72_{-1.32}^{+1.83}$ & $240.07_{-1.89}^{+2.84}$ \\
Recovered& $0.30_{-0.01}^{+0.08}$ & $-1.58_{-1.96}^{+2.66}$ & $-4.38_{-4.98}^{+2.94}$ & $23.43_{-10.56}^{+50.80}$ & $246.80_{-3.42}^{+10.89}$ \\
\hline
\end{tabular}
\end{center}
\end{table*}

The second example is a FLASH simulation with static gravity and radiative cooling of a cluster with an AGN in
the center (hereafter denoted as ``AGN feedback''); a simulation
snapshot was kindly provided to us by Marcus Br\"{u}ggen. The
simulated cluster, meant to mimic Hydra A, is described in
\citet{bruggen07} and \citet{simionescu09}; numerous plots of the
velocity field can also be found in \citet{vazza12}. Here, we briefly
summarize some properties. 
The box size was 1 Mpc and AMR resolution reached a peak of 0.5 kpc
in the center, and a maximum of (1,4,8) kpc outside (16,100,200) kpc
respectively. A bipolar jet 2 kpc in diameter with power $L_{\rm jet}
= 3 \times 10^{45} \, {\rm erg \, s^{-1}}$ was then introduced; for
the analyzed snapshot the bulk velocity along the jet is $\sim
1500-1800 \, {\rm km s^{-1}}$, and a $M\sim 1.3$ shock has been driven into the
surrounding ICM. The  
AGN was also given a bulk velocity of $\sim670 \, {\rm km \, s^{-1}}$ along the direction of
(-1,1,0) relative to the ambient ICM, to mimic the observed offset
between the shock center and the AGN in Hydra A. 
In Fig \ref{fig:agnmap}, we show a 1 Mpc size density and velocity field map on the y-z plane through the
  center. The size of the figure is
1 Mpc. The large bulk velocity along the x-direction has been
subtracted from the figure.
The outflows from the
AGN stir 
the gas in the central region ($\sim 300$ kpc in radius), 
while the ambient gas
is left relatively quiescent\footnote{The
small velocity dispersion ($\sim 45 \, {\rm km \, s^{-1}}$) of the quiescent region in this example come from the fact that apart from AGN outburst, it is a relaxed cluster which has not experienced any recent major mergers. Note, however, that the initial conditions come from cosmological GADGET SPH simulations where the small scale gas motions may not have been fully resolved.}. The velocity field is predominantly radial outside 100 kpc (associated with jet expansion and the running shock), while it is close to isotropic within 100 kpc, indicating that instabilities have efficiently isotropized and distributed the jet power.\footnote{Note, however, as also discussed by \citet{vazza12}, that these simulations are purely hydrodynamic, and magnetohydrodynamic (MHD) effects can strongly affect fluid instabilities and energy transfer from AGN bubbles to the ICM \citep{ruszkowski07,dursi08,oneill09}. For instance, 3D MHD simulations of bipolar jets by \citet{oneill10} find to the contrary that jet energy is not efficiently distributed/isotropized, remaining instead near the jet/cocoon boundary.} In Fig. \ref{fig:agn}, we plot the 
emission-weighted velocity PDF along the z-direction
inside an area of $1\times1~{\rm Mpc}^2$. The division between turbulent and quiescent gas shows
up in the velocity PDF as a double Gaussian distribution -- a narrow
Gaussian corresponding to the quiescent gas outside the core, and a broad
Gaussian corresponding to the turbulent gas in the center. This is an example of non-volume filling turbulence discussed in \S~\ref{subsec:others}. 
Note that we have pessimistically chosen a viewing direction in which
there are no bulk motions (similar to the ``width driven'' case of the
preceding example). For other viewing angles, the jet expansion drives
bulk motions which result in two clear peaks in the spectrum (similar
to the preceding ``separation driven'' case).  

The scales in this Hydra A example are so large that in this
particular instance, the velocity structure could be spatially
resolved by Astro-H. However, Hydra A is of course an extremely rare
and energetic outburst; for more typical jet luminosities of $L_{\rm
  jet} \sim  10^{44} \, {\rm erg \, s^{-1}}$, the turbulently stirred
region will be at least a factor of $\sim 30^{1/3}\sim 3$ smaller or
$\sim 100$ kpc in size, and hence barely resolved by Astro-H. In this
instance, mixture modeling will still be required to uncover the
filling fraction of turbulence. Also, as previously discussed, MHD
simulations show that motions are not efficiently isotropized and
distributed within the region of influence of the AGN, so in reality
there could be small scale intermittency in turbulence which would be
spatially unresolved, but detectable with mixture modeling.  

To approximate such situations, we analyze the spectrum with the
velocity PDF shown in Fig. \ref{fig:agn}, where the effects of line
blending, thermal broadening, 
instrument noise, and continuum emission have been included. This
is a clear example of the ``width driven'' scenario, with
$\sigma_{1}/\sigma_{2} = {0.70}$.  
We were unable to recover the velocity PDF from the mock spectrum with
$10^4$ photons. The estimated BIC values for the
single and double component models using the ``true values'' indeed
show that the single component model is preferred for $10^4$ photons
($BIC_{\rm double}-BIC_{\rm single}$=17).
However, with $10^5$ photons (which is for instance, possible for
Perseus; see Table \ref{tbl:clusters}), the two components could be
easily separated (in this case, $BIC_{\rm double}-BIC_{\rm single}$=-102). The
results are given in Table  
\ref{tbl:app2}. Again, here the ``true values'' are obtained by fitting the
PDF directly, by generating a Monte-Carlo sample of $10^4$
photons. The corresponding 2-D error contours
and marginalized posterior are shown in Fig. \ref{fig:agn_2D}. Note
the firm lower limit of $\sim 30\%$ to the quiescent component; a
clear detection that turbulence is not volume-filling. The velocity
dispersion and hence the energy density in the turbulent component are
also accurately recovered.

\section{Conclusions}
\label{sec:conclusions}

Gas motions can have profound influence on many physical processes in
the ICM, but thus far we have lacked a direct measurement of
turbulence in clusters. Upcoming X-ray missions--in particular
Astro-H--are poised to change that, by directly measuring turbulent
broadening of spectral lines. Thus far, most work has focussed on how
gas motions can alter the mean and width of X-ray emission lines from
galaxy clusters. However, the detailed shape of the line profile has
valuable information beyond these first two moments. Exploiting the
line shape (and thus the high spectral resolution of upcoming missions
such as Astro-H) can in many cases ameliorate poor angular resolution
in inferring the 3D velocity field. The main point of this paper is
that the line-of sight velocity PDF can often be meaningfully
decomposed into multiple distinct and physically significant
components. The 
separation is based on deviations of line profiles from a single
Gaussian shape, driven by either the difference in width (``width-driven'', WD) or
mean (``separation-driven'', SD) of the components. Such a mixture
decomposition yields {\it qualitatively} different results from a
single component fit, and the recovered mixture parameters have
physical significance. For instance, bulk flows and sloshing produce
components with offset means, while partial volume-filling turbulence
from AGN or galaxy stirring leads to components with different
widths. The offset between components allows us to measure gas bulk
motions and separate them from small-scale turbulence, while component
fractions and widths constrain the emission weighted volume and
turbulent energy density in each component. With the MCMC algorithm
and Fisher matrix techniques, we evaluate the 
prospects of using Gaussian mixture models to separate and constrain different velocity modes in galaxy clusters 
from the 6.7 keV Fe line complex. We found that with the $10^4$ photons (which is feasible for the $\sim 14$ nearest clusters; see Table \ref{tbl:clusters}), the components could be
constrained with $\sim 10$\% accuracy in WD cases, and $\sim 1$\%
accuracy in SD cases, in both toy models and simulations of clusters with cold fronts and AGN feedback respectively. Continuum emission degrades the constraints
in WD cases, while it has little impact on the SD cases. On the other hand, line blending appear to have little impact. We generally find
that Astro-H is effective in separating different components when
either the offset between the components or the width of one of the
components is larger than $\sim 200$ km/s. Using PDFs taken from
numerical simulations as ``basis'' functions, we find that reasonable deviations from Gaussianity in the mixture components do not affect our results. We also study error scalings and use information criteria to determine when a mixture model is preferred.

Many extensions of this method are possible. For instance: (i) It
would be interesting to compare the separation between bulk/turbulent
motions obtained from mock X-ray spectra by mixture modeling, with
algorithms for performing this separation for the full 3D velocity
field in numerical simulations (e.g., \citet{vazza12}), to see how
close the correspondence is. (ii) In this study, we have assumed that
due to Astro-H's poor spatial resolution, only line-of-sight
information about the velocity field is possible. In principle, it
should be possible also to obtain information about variation of the
velocity field in the plane of the sky. For nearby clusters such as
Perseus, it should be possible to examine the line shape as a function
of projected radial position to obtain a full 3D reconstruction of the
velocity field (a more detailed implementation of the suggestion by
\citet{zhuravleva12} to study the variation of line center and width
with projected radial position). It would be very interesting to study
the variation of mixture parameters as a function of position in
high-resolution simulations. Even for more distant clusters, a
coarse-grained tiling of the cluster should be possible. (iii) High
resolution X-ray imaging of cold-front clusters yield information
about density/temperature contact discontinuities in the plane of the
sky. This has already been used infer the presence of sloshing and
bulk motions, as well as physical properties of the ICM such as
viscosity and thermal conductivity. Combining information about the
density/temperature contact discontinuity in the plane of the sky with
the line of sight information obtained by mixture modeling could
enhance our understanding of gas sloshing in clusters, and give more
precise constraints on velocities. It would likewise be interesting to
employ mixture modeling on spectra of the violent merger clusters with
classic bow shocks.

More generally, mixture modeling of spectra should prove useful whenever there are good reasons to believe that there are multiple components to the thermal or velocity field, and/or the line profile shows significant deviations from Gaussianity. For instance, it might be fruitful to consider applications to the ISM (e.g., \citet{falgarone04,Lazarian2006}), or Ly$\alpha$ emission from galaxies (e.g., \citet{hansen06,dijkstra12}). 

\vspace{-0.5\baselineskip}

\section*{Acknowledgments}

We thank Marcus Br\"{u}ggen for kindly providing a simulation snapshot
of AGN feedback from \citet{bruggen07}, the authors of
\citet{Vazza2010} for making their simulation data publicly available,
and Brendon Kelly, Chris Reynolds, Franco Vazza, Sebastian Heinz and
Fanesca Young for 
helpful conversations or correspondence. We acknowledge NSF grant
AST0908480 for support. SPO thanks UCLA and KITP (supported in part by
the National Science Foundation under Grant No. NSF PHY05-51164) for
hospitality. We acknowledge the use of facilities at the Center for
Scientific Computing at the CNSI and MRL (supported by NSF MRSEC
(DMR-1121053) and NSF CNS-0960316).   


\bibliography{library,turb}

\appendix

\section[]{Gibbs Sampling MCMC Mixture Models}
\label{section:gibbs}

In this paper, we have adopted a ``poor man's'' approach to mixture modeling, in that we have binned the data, and dealt with the log-Poisson likelihood of the binned data. In principle, binning destroys information; however, in practice we have found that given the large number of data points, this information loss is negligible. We chose to adopt this approach since it is simpler and faster (each Monte Carlo simulation now requires operations over $\sim 50$ bins rather than $\sim 10^{4}$ data points; hence it is about $\sim 200$ times faster). In this Appendix, we describe a more statistically rigorous way of Bayesian parameter estimation with mixture models \citep{roeder97,gelman04,marin05,kelly07}, and show how it compares with this simpler method. We use the convention used in statistical literature of ``$\sim$'' to denote ``is distributed as'' or ``is drawn from'', rather than the usual ``is to rough order of magnitude'' in astronomical literature. 

We will perform MCMC samples of the full posterior distribution of the mixture model. This requires priors to be specified; formally, for Gaussian mixture models, an injudicious choice (such as a uniform prior) can lead to improper (non-integrable) posterior densities (see \citet{roeder97} for further discussion of this issue). A particularly convenient choice of priors are {\it conjugate priors}, in which the posterior distributions $p(\theta|x)$ are in the same family as the prior probability distribution $p(\theta)$. For instance, the Gaussian family is conjugate to itself (or self-conjugate) with respect to a Gaussian likelihood function: if the likelihood function is Gaussian, choosing a Gaussian prior over the mean will ensure that the posterior distribution is also Gaussian. Fortunately, all members of the exponential family have conjugate priors; for our mixture model, they are \citep{gelman04}:
\begin{eqnarray}
\label{eqn:prior_mix}
(\omega_{1},\ldots, \omega_{k}) &\sim& {\rm Dir}(a_{1},\ldots,a_{k}) \\
\label{eqn:prior_mean}
\mu_{j} &\sim&{\cal N}(\tilde{\mu}_{j},\tau_{j}^{2}) \\
\label{eqn:prior_var} 
\sigma_{j}^{2} &\sim& {\rm Inv-Gamma}(\alpha_{j},\beta_{j}). 
\end{eqnarray}

These prior distributions in turn require further parameters--known as {\it hyper-parameters}--to be specified. In the absence of prior information, techniques exist to allow these hyper-parameters to become additional parameters in the statistical fit, and thus be determined by the data itself \citep{roeder97}. However, given that we do have guidance from both theory and other observations (e.g. X-ray imaging; morphology; temperature and density profiles) about the expect thermal and turbulent broadening of the spectrum, it is reasonable to adopt fixed priors. Here we adopt uninformative priors, but describe the choice of priors in some detail in case more informative priors are desired. 
\begin{itemize} 
{\item {\bf Mixture weights $\omega_{j}$}. The prior for the mixture weights, $(\omega_{1},\ldots, \omega_{k}) \sim {\rm Dir}(a_{1},\ldots,a_{k})$ is a Dirichlet distribution, which is conjugate to the multinomial distribution, the following sense: given a prior ${\rm Dir}(a_{1},\ldots,a_{k})$, then if $(b_{1},\ldots,b_{k})$ is the number of occurrences of each event $i$ in a sample of $n$ events, the posterior is ${\rm Dir}({\bf a}+{\bf b})$. If there is no prior information to favor one component over another, then it is common practice to set all members of the prior $a_{i}$ to a common value $a$, known as the concentration parameter; $a \ll 1$ favors concentration amongst a few components, whilst $a \gg 1$ favors almost equal dispersion amongst all components. Here, we adopt a ${\rm Dirichlet}(1,\ldots,1)$ prior, which is equivalent to a uniform prior on ${\bf \omega}$, under the constraint that $\sum_{j=1}^{k} \omega_{j} = 1$.}
{\item {\bf Line centers $\mu_{j}$}. We adopt a normal prior for the line centers, $\mu_{j} \sim {\cal N}(\tilde{\mu}_{j},\tau_{j}^{2})$, where $\tilde{\mu}_{j}$ are the known line centers in the absence of turbulent motions. We set a very weak prior $\tilde{\mu}_{j}=\mu$, where $\mu$ is the mean energy of all photons, and $\tau_{j}=(c_{s}/(\sqrt{3}c) E_{o}$, where $c_{s}/\sqrt{3}$ corresponds to typical line of sight velocities of transonic bulk motions, and $E_{o}$ is the line center energy.}
{\item {\bf Line widths $\sigma^{2}_{j}$}. The prior is the inverse Gamma distribution, $\sigma_{j}^{2} \sim {\rm Inv-Gamma}(\alpha_{j},\beta_{j})$; the inverse Gamma distribution has the desirable property that it is bounded below at zero, and is defined such that $1/\sigma_{j}^{2}$ obeys a Gamma distribution. Two constraints are needed to determine the two parameters of the distribution, $(\alpha_{j},\beta_{j})$. Since the inverse-Gamma distribution is highly asymmetric, with a long tail to high values of $\sigma_{j}^{2}$, we do not use estimates for the first two moments of the distribution to set $(\alpha_{j},\beta_{j})$. Instead, we set the mode of the distribution to $\sigma_{\rm mode}^{2} = \sigma_{\rm low}^{2}=\beta_{i}/(\alpha_{i}+1)$, where $\sigma_{\rm low}^{2}=\sigma^{2}_{\rm therm} + \sigma^{2}_{\rm instrum}$ is the line width in the absence of turbulence; i.e., due to thermal and instrumental broadening alone. The asymmetric nature of the Inv-Gamma distribution generally implies that $P(\sigma^{2} < \sigma_{\rm mode}^{2}) \sim 0.1-0.2$. We require that $P(\sigma^{2} > \sigma_{\rm high}^{2}) = 0.1$, where $\sigma^{2}_{\rm high}=\sigma^{2}_{\rm low} + v_{\parallel, {\rm max}}^{2}$, and $v_{\parallel, {\rm max}}^{2}$ is such that $U_{\rm turb} \sim 0.5 U_{\rm therm}$ for isotropic turbulence. Together, these two requirements allow $\alpha, \beta$ to be determined, and result in $\sim 70-80\%$ of samples drawn from the prior to lie between $\sigma^{2}_{\rm low}$ and $\sigma^{2}_{\rm high}$. This fairly loose prior allows $\sigma^{2}_{j}$ to be mostly driven by the data. A similar procedure can be used for more informative priors.}
\end{itemize} 

Our goal is to calculate the posterior of the model parameters, given the prior distribution and the data, and draw samples from this posterior in a numerically tractable manner. Equation (\ref{eqn:mixture}) in \S\ref{sec:methodology} is the probability of drawing a data point with value $x$, given the model parameters $\theta$. Since data points are independent, the likelihood of a data set is the product of the probabilities for each data point: 
\begin{equation}
{\cal L}(x|\theta) = \prod_{i} f(x_{i}|\theta).  
\end{equation}
The posterior probability of the model parameters $\theta$, given the data $x$ and priors for the model parameters $p(\theta|\tilde{\theta})$, where $\tilde{\theta}$ are the known hyper-parameters, is then simply: 
\begin{equation}
P(\theta|x,\tilde{\theta}) \propto {\cal L}(x|\theta) p(\theta|\tilde{\theta}). 
\end{equation} 
In numerical work, we will usually work with the logarithm of the posterior, rather than the posterior itself, to avoid problems with underflow and rounding to zero. However, from equation (\ref{eqn:mixture}), evaluating the posterior involves the sum of Gaussians, each of which may be so small as to underflow when reconstructed from their logarithms. One must therefore use the {\it log-sum-exp} formula (e.g., see \citet{press07}):  
\begin{equation} 
{\rm log}\left(\sum_{i} {\rm exp}(z_{i}) \right) = z_{\rm max} + {\rm log}\left(\sum_{i} {\rm exp}(z_{i}-z_{\rm max}) \right),
\end{equation} 
where $z_{i}$ are the logarithms of small quantities and $z_{\rm max}$ is their maximum. This guarantees that at least one exponentiation won't underflow, and any that do could have been neglected anyhow. 

We now draw Markov-Chain Monte Carlo (MCMC) samples from the posterior distribution. The Metropolis-Hastings algorithm for drawing MCMC samples is probably the one most familiar to readers; for instance, it is the heart of the widely used software package COSMOMC \citep{Lewis2002}, which we also used previously. Here we use instead Gibbs sampling, which is the most commonly used approach in Bayesian mixture estimate (e.g., \citet{marin05}, and references therein); it is in fact a special case of the Metropolis-Hastings algorithm \citep{gilks96}, even though historically it was developed separately. In Gibbs sampling, one draws each parameter $\theta_{i}$ from its full conditional distribution, which is obtained by holding all components of $\theta$ constant except for $\theta_{i}$, and sampling from the posterior as a function of $\theta_{i}$ alone. It can be shown that (unlike the Metropolis-Hastings algorithm) the acceptance probability is one in this case, i.e., we can {\it always} accept a sample drawn from the conditional distribution $P(\theta_{i}|\theta^{-})$ (where $\theta^{-}$ denotes ``values of all parameters except one''). To draw a sample, one simply cycles through each component of $\theta$ in turn. Note that, unlike the Metropolis-Hastings method, each component of $\theta$ gets reset to a value completely independent of its previous value (although $\theta^{-}$ does depend on previous values; thus, we still have a Markov chain). Thus, in principle larger steps in parameter space are possible than with MH sampling\footnote{However, despite this, Gibbs sampling does not always enjoy good convergence properties, and one has to be wary of getting trapped in local minima \citep{gilks96}.}, since with the latter, large multi-variate steps will have low acceptance probabilities and almost certainly be rejected. Despite these obvious advantages, a major disadvantage of Gibbs sampling is that it requires the full conditional distribution to be correctly normalized (unlike MH sampling, where an unnormalized posterior is sufficient), and thus the numerically expensive computation of normalizing constants for each $\theta^{-}$ (as well as a practical means of drawing $\theta_{i}$ from each conditional distribution). Fortunately, for Gaussian mixture models, the use of data augmentation and conjugate priors allows for analytic conditional distributions, for which there are well-known random number generators. 

We adopt a data augmentation approach to the ``missing data'' problem previously mentioned: the particular mixture to which a data point belongs is unknown, obviating a straightforward calculation of model parameters. We can treat this unknown membership by assigning to each observation $x_{i}$ a new variable $z_{i} \in {1,2,\ldots,k}$ which indicates which mixture the data point belongs to. Given the augmented membership variable $z_{i}$, equation (\ref{eqn:mixture}) then reduces to: 
\begin{equation}
f(x_{i}|z_{i}=j,\theta) \sim f_{j}(x_{i}|\mu_{j},\sigma^{2}_{j})
\end{equation}
where $f_{j}(x|\mu_{j},\sigma^{2}_{j})$ is the $j$th normal mixture component. Of course, we do not know the $z_{i}$, but we know their probability distribution: since the probability that a data point $x_{i}$ belongs to a mixture  $j$ is $p_{j} \propto w_{j} f_{j}(x_{i}|\mu_{j},\sigma^{2}_{j})$ (normalized such that $\sum_{j} p_{j}=1$), the $z_{i}$ belong to a multinomial distribution $z_{i} \sim {\cal M}_{j}(1;p_{1},\ldots,p_{k})$. 

With this membership variable $z_{i}$ and the conjugate priors in equations (\ref{eqn:prior_mix}),(\ref{eqn:prior_mean}),(\ref{eqn:prior_var}), we can compute the conditional distributions, which all take the functional form of their conjugate priors. There are well-established algorithms for drawing random numbers from all of these analytic distributions. The mixing fractions $\omega$ are Dirichlet distributed: 
\begin{equation}
(\omega_{1},\ldots, \omega_{k}) \sim {\rm Dir}(a_{1}+n_{1},\ldots,a_{k}+n_{k}) 
\end{equation} 
where $n_{j}$ are the number of data points in mixture $j$, obtained by summing over the indicator variables $z_{i}$. We can immediately see that the hyper-parameters we have chosen $a_{1},\ldots,a_{k}=1$ will have little influence for a large sample. The means $\mu_{j}$ are Gaussian distributed: 
\begin{equation}
\mu_{j} \sim {\cal N}\left( \frac{\frac{\tilde{\mu}_{j}}{\tau_{j}^{2}} + \frac{\sum_{i|z_{i}=j} x_{i}}{\sigma_{j}^{2}}}{\frac{n_{j}}{\sigma_{j}^{2}} + \frac{1}{\tau_{j}^{2}}}, \left( \frac{n_{j}}{\sigma_{j}^{2}} + \frac{1}{\tau_{j}^{2}} \right)^{-1} \right) 
\end{equation} 
where the sum runs over all data points $x_{i}$ which belong to mixture $j$, as indicated by the membership variable $z_{i}=j$. The variances are drawn from an inverse-Gamma distribution: 
\begin{equation} 
\sigma^{2}_{j} \sim {\rm Inv-Gamma}\left(\alpha_{j} + \frac{n_{j}}{2}, \beta_{j} + \frac{1}{2} \sum_{i|z_{i}=j} (x_{i} - \mu_{j})^{2} \right). 
\end{equation}
Finally, at every iteration we have to draw a new set of indicator variables $z_{i}$ from a multinomial distribution: 
\begin{equation}
z_{i} \sim {\cal M}(1;p_{1},\ldots,p_{k})
\end{equation} 
and $p_{j}(x_{i}) \propto w_{j} f_{j}(x_{i}|\mu_{j},\sigma^{2}_{j})$ (normalized such that $\sum_{j} p_{j}=1$). A set of each of these random draws then comprises a single Monte-Carlo sample from the posterior distribution. In our calculations, we find that computing time is heavily dominated by random draws from the multinomial random number generator. We therefore parallelized the multinomial random number generator, and found that this allowed a significant (close to linear) reduction in computation time\footnote{An alternative would be to run multiple MCMC streams in parallel. However, due to the long burn in period for certain regions of parameter space, this strategy is less efficient.}.  

As before, we deal with the ``label-switching'' problem by demanding that $\mu_{1} < \mu_{2}$. Note that this is still an area of active research (e.g., see \citet{jasra05}); as yet there is no consensus solution in the statistical community. Alternatively, we find that if we enforce asymmetric priors, this is usually sufficient to break the symmetry in the modes of the posterior and single out a single one.

We have experimented with running this Gibbs sampling MCMC and the Cash-C Metropolis-Hastings MCMC described in the text, and found that they give similar results. For instance, it gives almost exactly the same results for the single line SD case given in Table \ref{tbl:para_sd}, and similar results for the WD case given in Table \ref{tbl:para_wd}. Interestingly, often the trace plots of the Gibbs sampling MCMC method for the WD case show that it has somewhat poorer mixing and convergence properties---indicating that it is more badly affected by parameter degeneracies. The greater robustness of the method we used in this paper stems from the use of covariance matrix information to improve the proposal distribution in COSMOMC \citep{Lewis2002}; a similar means of orthogonalizing parameters should be implemented here to ensure robustness. Overall, there is no reason to use full mixture modeling for the purposes of this paper. However, the algorithm described in this Appendix should be preferred when there are fewer data points, and binning is deprecated. 

\appendix

\label{lastpage}
\end{document}